\documentclass[manuscript,screen]{acmart}

\AtBeginDocument{%
  \providecommand\BibTeX{{%
    \normalfont B\kern-0.5em{\scshape i\kern-0.25em b}\kern-0.8em\TeX}}}

\setcopyright{none}
\copyrightyear{2024}
\acmYear{2024}
\acmDOI{XXXXXXX.XXXXXXX}

\acmJournal{TOIS}
\acmVolume{XX}
\acmNumber{X}
\acmArticle{XXX}
\acmISBN{978-1-4503-XXXX-X/18/06}

\usepackage{algorithm}
 \usepackage{amssymb}    
\usepackage{bm}         
\usepackage{amsmath}    
\usepackage{booktabs}   
\usepackage{tabularx}   
\usepackage{siunitx}    
\usepackage{svg}        
\usepackage{makecell}   
\usepackage{appendix}   
\usepackage{float}      
\usepackage{xcolor}   
\usepackage{hyperref}      
\usepackage{graphicx}
\usepackage{subfig}
\usepackage{ulem}
\usepackage{multirow}
\usepackage{subfiles} 
\usepackage{algpseudocode}

\newcommand{\Model}{\texttt{HypeMed}}
\newcommand{\MedRep}{MedRep}
\newcommand{\MedRec}{SimMR}

\newif\ifshowrev
\showrevfalse

\ifshowrev
  \newcommand{\rev}[1]{\textcolor{blue}{#1}}
\else
  \newcommand{\rev}[1]{#1}
\fi

\begin{document}
\title{HypeMed: Enhancing Medication Recommendations with Hypergraph-Based Patient Relationships}

\author{Xiangxu Zhang}
\affiliation{%
  \institution{Gaoling School of Artificial Intelligence, Renmin University of China}
  \city{Beijing}
  \country{China}
  \postcode{100872}
}
\email{xansar@ruc.edu.cn}

\author{Xiao Zhou}
\authornote{Corresponding author.}
\affiliation{%
  \institution{Gaoling School of Artificial Intelligence, Renmin University of China}
  \city{Beijing}\country{China}}
\additionalaffiliation{%
  \institution{Beijing Key Laboratory of Research on Large Models and Intelligent Governance}
  \city{Beijing}\country{China}}
\additionalaffiliation{%
  \institution{Engineering Research Center of Next-Generation Intelligent Search and Recommendation, MOE}
  \city{Beijing}\country{China}}
\email{xiaozhou@ruc.edu.cn}

\author{Hongteng Xu}
\authornotemark[2] 
\authornotemark[3] 
\affiliation{%
  \institution{Gaoling School of Artificial Intelligence, Renmin University of China}
  \city{Beijing}
  \country{China}
  \postcode{100872}
}
\email{hongtengxu313@gmail.com}

\author{Jianxun Lian}
\affiliation{%
  \institution{Microsoft Research Asia}
  \city{Beijing}
  \country{China}
  \postcode{100080}
}
\email{jianxun.lian@outlook.com}

\renewcommand{\shortauthors}{Zhang, et al.}

\begin{abstract}

Medication recommendation aims to generate safe and effective medication sets from health records. However, accurately recommending medications hinges on inferring a patient's latent clinical condition from sparse and noisy observations, which requires both (i) preserving the visit-level combinatorial semantics of co-occurring diagnoses/procedures and (ii) leveraging informative historical references through effective, visit-conditioned retrieval. 
Most existing methods fall short in one of these aspects: graph-based modeling often fragments higher-order intra-visit patterns into pairwise relations, while inter-visit augmentation methods commonly exhibit an imbalance between learning a globally stable representation space and performing dynamic retrieval within it.
To address these limitations, this paper proposes \Model, a two-stage hypergraph-based framework unifying intra-visit coherence modeling and inter-visit augmentation.
\Model\ consists of two components: \MedRep\ for representation pre-training and \MedRec\ for similarity-enhanced recommendation.
In the first stage, \MedRep\ encodes clinical visits as hyperedges via knowledge-aware contrastive pre-training, creating a globally consistent, retrieval-friendly embedding space.
In the second stage, \MedRec\ performs dynamic retrieval within this space, fusing retrieved references with the patient's longitudinal data to refine medication prediction.
Evaluation on real-world benchmarks shows that \Model\ outperforms state-of-the-art baselines in both recommendation precision and DDI reduction, simultaneously enhancing the effectiveness and safety of clinical decision support.
The implementation is publicly available at \href{https://github.com/xansar/HypeMed}{\textcolor{blue}{https://github.com/xansar/HypeMed}}.
\end{abstract}





\maketitle

    \begin{figure}[htb]
        \centering
        \includegraphics[width=\textwidth, keepaspectratio]{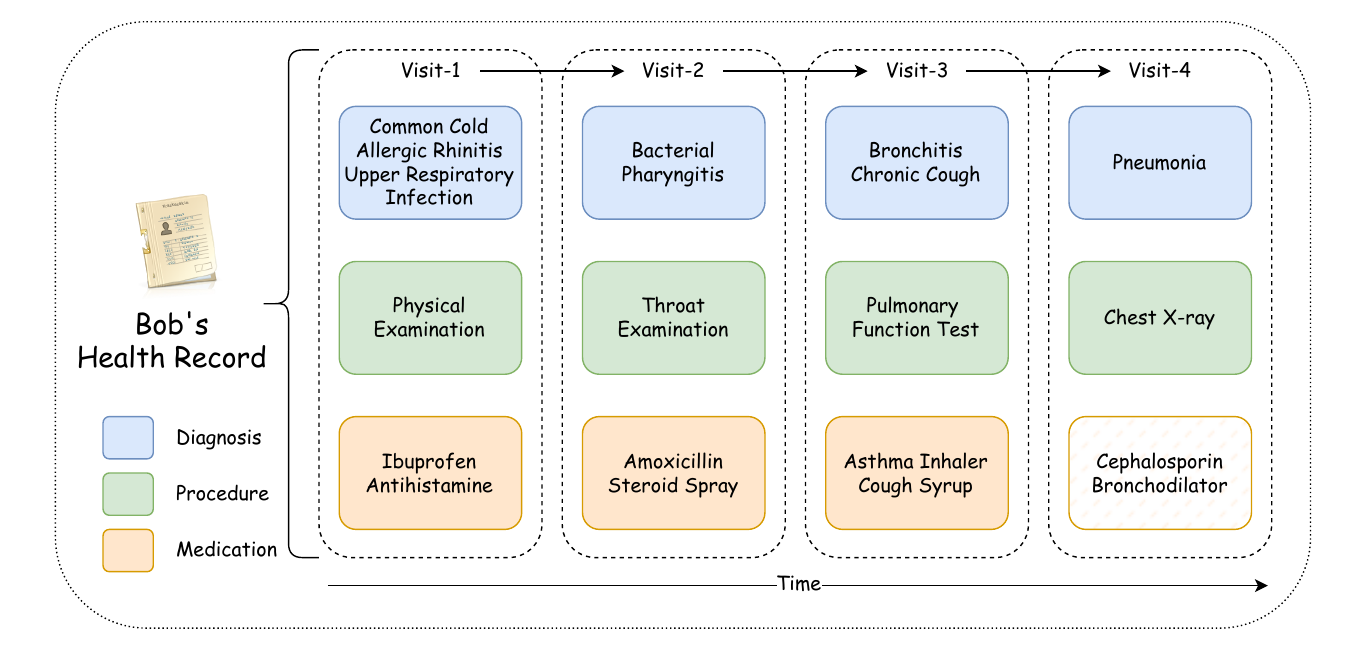}\\
        \caption{
            Overview of Bob's EHR spanning four medical visits. Each visit includes diagnosis, procedure, and medication codes, illustrating the progression of related respiratory conditions over time. Note that the medication section in Visit-4 is left unfilled, indicating it as the target for medication recommendation.
        }
        \label{fig:ehr}
    \end{figure}

\section{Introduction}
\rev{Healthcare remains a fundamental societal challenge~\cite{braveman2014social,blumenthal2020covid,li2025r2med}, intensified by rapid urbanization~\cite{he2025motifgpl,li2025fap,zhang2026social,yong2026intelli,li2025spatio,hong2026wed} and recurring public health crises~\cite{hong2025lost,zhu2025hypertension}. As a key application of Artificial Intelligence (AI)~\cite{zhu2025mohobench,guo2025counterfactual,yong2025motivebench,yongthink} in healthcare~\cite{jiang2017artificial,saraswat2022explainable}, medication recommendation systems~\cite{SafeDrug, COGNet, GAMENet, LEAP, RETAIN} aim to generate safe and effective medication combinations from patients' Electronic Health Records (EHRs)~\cite{menachemi2011benefits,evans2016electronic}. EHRs provide a longitudinal trace of diagnoses, procedures, and medications across visits (Fig.~\ref{fig:ehr}). While EHRs offer a wealth of information, the inherent complexity and high dimensionality of such data make it difficult to capture the complete clinical picture.}

\rev{Despite steady progress~\cite{SafeDrug, COGNet, yang2023molerec, VITA, promise, DAPSNet}, the central difficulty remains unchanged: \emph{medication recommendations hinge on accurately inferring a patient's latent clinical condition from sparse, noisy observations}. The observed medical codes provide only indirect, discrete evidence of an underlying physiological state, making condition reconstruction inherently ambiguous. To approximate the true patient status, a model must integrate evidence not only from the current observations but also from their \emph{relationships} in EHRs.
In this paper, we use \emph{patient relationships} to refer to two complementary forms of clinical evidence: (i) \emph{intra-visit} set-level co-occurrence among diagnoses/procedures/medications, which reflects coherent syndromic patterns, and (ii) \emph{inter-visit} connections that link the current visit to relevant visits in the patient trajectory, including prior visits of the same patient and clinically similar visits from other patients.
However, existing methods often fail to unify these two forms of evidence, yielding a fragmented and superficial estimate of the underlying condition.}

\rev{\textbf{From the intra-visit perspective}, a dominant bottleneck is \emph{semantic fragmentation}. Most approaches~\cite{GAMENet, SafeDrug, COGNet, DGCL} model co-occurrence with simple graphs, breaking inherently high-order clinical patterns into pairwise edges and thus diluting visit-level combinatorial semantics. For instance, Charcot's triad---fever, right-upper-quadrant pain, and jaundice---is individually non-specific, but their \emph{co-occurrence} is highly suggestive of acute cholangitis and warrants urgent care~\cite{frossard2011charcot, rumsey2017diagnostic}. This motivates modeling a visit as a set-level interaction, rather than an accumulation of pairwise relations.
\rev{\textbf{From the inter-visit perspective}, the key is to \emph{retrieve the right historical references} that can effectively complement the current visit for decision making~\cite{brooks1991role,solomon2006case,brown2016patient}. Existing methods often exhibit an imbalance between static representation learning and dynamic retrieval. Representation-centric approaches (e.g., DGCL~\cite{DGCL}, PROMISE~\cite{promise}) mainly focus on learning a globally stable and static embedding space by aggregating historical/external signals, yet they typically lack an explicit mechanism to perform visit-conditioned retrieval tailored to the current context. In contrast, retrieval-centric approaches (e.g., GAMENet~\cite{GAMENet}, COGNet~\cite{COGNet}, VITA~\cite{VITA}, DAPSNet~\cite{DAPSNet}) emphasize on-the-fly retrieval in the embedding space to augment the current input, but often rely on representations that are not explicitly optimized for retrieval, making the retrieved neighbors less semantically aligned with the current clinical context. This imbalance motivates a unified design that jointly shapes a retrieval-friendly representation space and performs context-aware retrieval within it.}}

\rev{To address these limitations, we propose \Model, a hypergraph-based framework that reconstructs latent clinical conditions by unifying intra-visit combinatorial interaction modeling with inter-visit reference augmentation, thereby capturing patient relationships both within and across visits. To mitigate \emph{semantic fragmentation}, we model each clinical visit as a hyperedge, so that the representation is learned from the set-level co-occurrence of diagnoses and procedures rather than decomposed pairwise links. Based on this structure, we introduce \textbf{Medical Entity Relevance Representation (\MedRep)}, a knowledge-aware hypergraph contrastive pre-training stage that encodes high-order interactions into a globally consistent embedding space. On top of the MedRep space, to alleviate the representation--retrieval imbalance, we further propose \textbf{Similar Visit Enhanced Medication Recommendation (\MedRec)}, which performs visit-conditioned dynamic retrieval directly in this hyperedge-aware metric space. This coupled design makes the retrieved neighbors better aligned with the current visit context, enabling \Model\ to incorporate informative historical references and refine the estimated condition for medication recommendation.
Extensive experiments on real-world benchmarks (MIMIC-III/IV and eICU) demonstrate that \Model\ consistently outperforms state-of-the-art baselines in recommendation accuracy while reducing drug–drug interaction (DDI) rates, thereby achieving a superior balance of clinical effectiveness and medication safety. Our main contributions are summarized as follows:}

\rev{\begin{itemize}
    \item We address semantic fragmentation in intra-visit modeling by representing each visit as a hyperedge and proposing a hypergraph-based contrastive pre-training module (\MedRep) to encode knowledge-aware entity representations that capture combinatorial semantics.
    \item We alleviate the representation--retrieval imbalance in inter-visit augmentation by designing a similar visit enhanced medication recommendation module (\MedRec) that performs visit-conditioned dynamic retrieval directly in the \MedRep-optimized embedding space for context-aligned reference aggregation.
    \item Extensive experiments on public MIMIC-III/IV and eICU datasets demonstrate that \Model\ consistently improves recommendation accuracy while reducing the drug--drug interaction (DDI) rate, validating both effectiveness and safety.
\end{itemize}}

    \section{Related Work}

    \subsection{Medication Recommendation}
    Medication recommendation is an important subdomain of recommender systems~\cite{ma2024tail,guo2025sorex,zhou2025tricolore,li2025leave,guo2026not}.
    \rev{Improving medication recommendation from EHRs hinges on two complementary capabilities: modeling \emph{intra-visit} clinical context from the current observations, and leveraging \emph{inter-visit} references from the patient trajectory and clinically similar visits from other patients.}
    
    \rev{\noindent\textbf{Intra-visit reasoning with relational structures.}
    A representative line of work leverages structured relational modeling---often via graph neural networks (GNNs)~\cite{GNN}---to capture concurrence among diagnoses, procedures, and medications, and to inject domain knowledge such as DDIs for safety-aware prediction~\cite{GAMENet, SafeDrug, COGNet, DGCL, Carmen, bhoi2021personalizing, zhang2023knowledge}. 
    While effective, these methods predominantly operate on \emph{pairwise} edges, which can fragment the visit-level context when the clinical meaning arises from the \emph{joint} presence of multiple entities.}
    
    \rev{\noindent\textbf{Inter-visit augmentation via similarity and retrieval.}
    Early studies mainly adopted \emph{instance-based} models that recommend medications from a single visit, without explicitly leveraging inter-visit context~\cite{LEAP, SMR}. 
    Later, \emph{longitudinal} models incorporated a patient's historical visits to capture temporal dependencies and improve prediction~\cite{RETAIN, DMNC, DrugRec, Carmen, DGCL, COGNet}. 
    More recently, \emph{external} augmentation extends the context beyond an individual patient by retrieving similar visits from the broader cohort, yet existing designs often emphasize only one side of the problem. PROMISE~\cite{promise} mainly focuses on learning a globally stable, trajectory-level representation (e.g., via DTW-based aggregation), but provides limited visit-conditioned, on-the-fly retrieval for the current context. In contrast, DAPSNet~\cite{DAPSNet} emphasizes dynamic retrieval at prediction time to augment the current visit, but relies on representations that are not explicitly optimized for retrieval, which may weaken the quality of retrieved references.
    However, these retrieval-based approaches often decouple similarity estimation from representation learning, which can induce a mismatch between the embedding space and the retrieval objective, ultimately limiting the utility of retrieved references.}

    \rev{\noindent\textbf{Our approach.}
    To fill these gaps, \Model\ unifies intra-visit coherence modeling with inter-visit reference augmentation in a single hypergraph framework. \MedRep\ represents each visit as a hyperedge and performs knowledge-aware hypergraph contrastive pre-training to encode high-order interactions into a globally consistent, retrieval-friendly embedding space, mitigating semantic fragmentation. Building upon this space, \MedRec\ performs visit-conditioned dynamic retrieval and integrates the retrieved references with the patient trajectory to refine latent condition estimation and improve medication recommendation.}

    \subsection{Hypergraph Contrastive Learning in Recommendation}
    Hypergraph contrastive learning~\cite{TriCL} has recently attracted increasing attention as an effective strategy to mitigate data sparsity and enhance representation robustness in recommender systems. 
    For instance, in the context of session-based recommendation, 
    DHCN~\cite{DHCN} applies contrastive learning to align session representations derived from hypergraphs and their corresponding line graphs. 
    MHCN~\cite{MHCN} extends this idea to social recommendation by contrasting local and global node representations, while 
    HCCF~\cite{HCCF} integrates collaborative filtering with contrastive learning to alleviate over-smoothing and sparsity issues. 
    Collectively, these methods demonstrate the potential of combining hypergraph structures with contrastive objectives to capture richer relational semantics. 
    However, their applications remain largely confined to general recommendation scenarios. 
    This paper adapts and extends hypergraph contrastive learning to medication recommendation, \rev{enabling set-level intra-visit modeling and inter-visit reference augmentation in a unified framework.}

    \section{Preliminaries}
    This section introduces the notations and definitions used in this paper and outlines the process of constructing the hypergraph. The main symbols are summarized in Tab.~\ref{tb:notation}.
    \begin{table}[h]
        \caption{{Summary of main notations used in this paper.}}
        \label{tb:notation}
        \begin{tabularx}{0.9\textwidth}{p{0.18\textwidth}X}
            \Xhline{1pt}
            \textbf{Notation} & \textbf{Definition} \\
            \Xhline{0.5pt}
            $\mathcal{R}$ & EHR database containing all patients’ visit sequences \\
            $\mathcal{S}^{(i)}_t$ & The $t$-th visit of patient $i$ (patient index $i$ is omitted when clear) \\
            $\mathcal{D}, \mathcal{P}, \mathcal{M}$ & Sets of all diagnosis, procedure, and medication codes \\
            $\mathbf{d}, \mathbf{p}, \mathbf{m}$ & Multi-hot vectors for diagnoses, procedures, and medications\\
            $\mathbf{v}^{\mathcal{D}}$, $\mathbf{v}^{\mathcal{P}}$, and $\mathbf{v}^{\mathcal{M}}$ & Visit representations derived from diagnoses, procedures, and medications \\
            $\mathcal{H}_\mathcal{X}=(\mathcal{N}_\mathcal{X},\mathcal{E}_\mathcal{X})$ & Hypergraph for domain $X\!\in\!\{\mathcal{D},\mathcal{P},\mathcal{M}\}$ \\
            $\mathbf{Z}^\mathcal{X}, \mathbf{U}^\mathcal{X}$ & Node and hyperedge embeddings learned from $\mathcal{H}_\mathcal{X}$ \\
            $\mathbf{h}_t$ & Health status representation of visit $t$ (derived from diagnoses and procedures) \\
            $\mathbf{v}_{hist}$, $\mathbf{v}_{sim}$ & Representations from the historical-visit and similar-visit channels \\
            $\mathbf{v}_t$ & Final fused visit representation \\
            $\mathbf{y}_t$ & Predicted probability vector of medications for visit $t$ \\
            $\mathbf{A}$ & Drug–drug interaction (DDI) adjacency matrix \\
            \Xhline{1pt}
        \end{tabularx}
    \end{table}

    \subsection{Notations and Definitions}
    \subsubsection{Definition of EHR Data.}
    The fundamental unit of EHR data is the patient, and each patient’s EHR consists of multiple visits. 
    Let $\mathcal{R} = \{\mathcal{S}^{(i)}\}_{i=1}^{N}$ represent the EHR data of $N$ patients.
    Each patient $i$ has a patient trajectory, represented as an ordered sequence of visits $\mathcal{S}^{(i)} = \langle \mathcal{S}_1, \mathcal{S}_2, \ldots, \mathcal{S}_{|\mathcal{S}^{(i)}|} \rangle$,
    where each visit $\mathcal{S}_t = \langle \mathbf{d}_t, \mathbf{p}_t, \mathbf{m}_t \rangle$ 
    contains diagnosis, procedure, and medication multi-hot vectors,
    with $\mathbf{d}_t \!\in\! \{0,1\}^{|\mathcal{D}|}$, 
    $\mathbf{p}_t \!\in\! \{0,1\}^{|\mathcal{P}|}$, 
    and $\mathbf{m}_t \!\in\! \{0,1\}^{|\mathcal{M}|}$.
    Given the current diagnosis $\mathbf{d}_t$, procedure $\mathbf{p}_t$, and the EHR history 
    $\langle \mathcal{S}_1, \mathcal{S}_2, \ldots, \mathcal{S}_{t-1} \rangle$, 
    \Model~aims to predict the medication set $\mathbf{m}_t$ for the current visit.
    Patient indices $(i)$ and domain subscripts are omitted when clear from context.

    \subsubsection{Definition of Hypergraph.}
\begin{figure}[htb]
        \centering
        \includegraphics[width=0.5\textwidth, keepaspectratio]{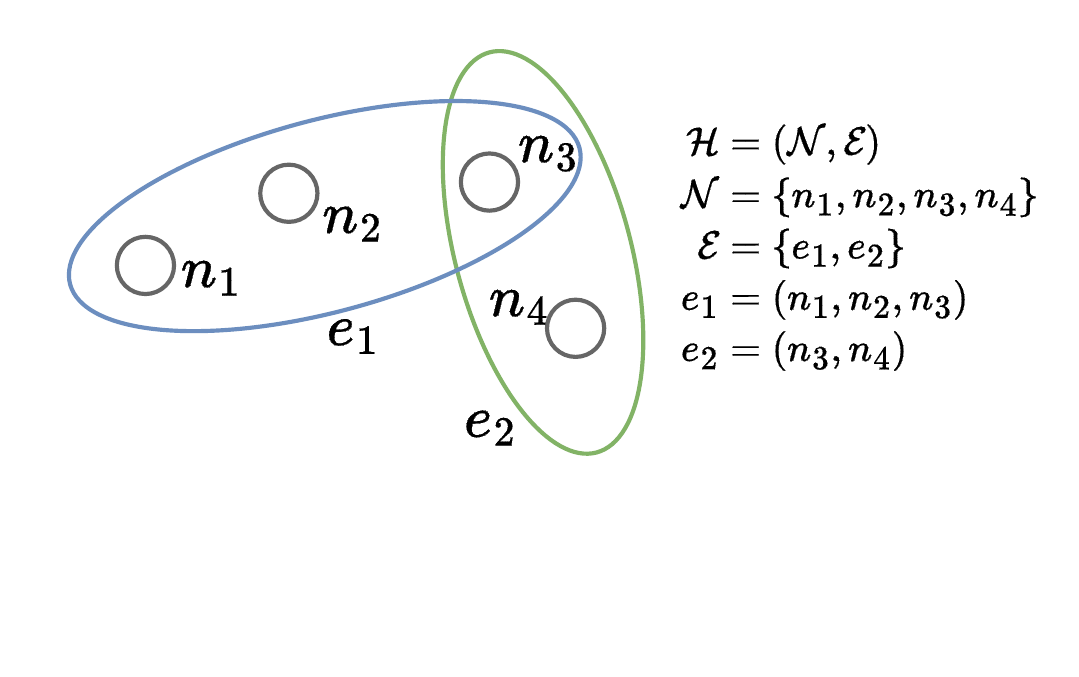}\\
        \caption{
            {An example of a hypergraph with four nodes and two hyperedges: $e_1$ connects nodes $n_1$, $n_2$, and $n_3$, while $e_2$ connects nodes $n_3$ and $n_4$.}
        }
        \label{fig:hypergraph}
    \end{figure}

    A hypergraph is denoted as $\mathcal{H}=(\mathcal{N},\mathcal{E})$, where $\mathcal{N}$ and $\mathcal{E}$ are the sets of nodes and hyperedges, respectively.
    Fig.~\ref{fig:hypergraph} illustrates an example hypergraph with four nodes and two hyperedges, where $\mathcal{N} = \{n_i\}_{i=1}^4$ and $\mathcal{E} = \{e_i\}_{i=1}^2$. Each hyperedge can connect more than two nodes; for example, $e_1$ connects $n_1$, $n_2$, and $n_3$.
    $\mathcal{N}(n_i)$ denotes the set of hyperedges incident to node $n_i$, and $\mathcal{N}(e_i)$ represents the set of nodes contained in hyperedge $e_i$.

    \begin{algorithm}
    \caption{Hypergraph Construction for Diagnosis, Procedure, and Medication}
    \label{alg:hypergraph}
    \begin{algorithmic}[1]
    \Function{ConstructHypergraphs}{$\{\mathcal{H}_{X}\}_{X\in\{\mathcal{D},\mathcal{P},\mathcal{M}\}}, \mathcal{R}$}
        \State $N_{\mathcal{D}}, N_{\mathcal{P}}, N_{\mathcal{M}} \gets \emptyset$ \Comment{Initialize node sets}
        \State $E_{\mathcal{D}}, E_{\mathcal{P}}, E_{\mathcal{M}} \gets \emptyset$ \Comment{Initialize hyperedge sets}
        \State \Call{AddEntitiesToHypergraph}{$\mathcal{D}, N_{\mathcal{D}}, E_{\mathcal{D}}, \mathcal{R}$}
        {\State \Call{AddEntitiesToHypergraph}{$\mathcal{P}, N_{\mathcal{P}}, E_{\mathcal{P}}, \mathcal{R}$}}
        \State \Call{AddEntitiesToHypergraph}{$\mathcal{M}, N_{\mathcal{M}}, E_{\mathcal{M}}, \mathcal{R}$}
        \State \textbf{return} $(N_{\mathcal{D}}, E_{\mathcal{D}}), (N_{\mathcal{P}}, E_{\mathcal{P}}), (N_{\mathcal{M}}, E_{\mathcal{M}})$
    \EndFunction
    \vspace{5mm}
    \Procedure{AddEntitiesToHypergraph}{$\mathcal{X}, N_{\mathcal{X}}, E_{\mathcal{X}}, \mathcal{R}$}
        \ForAll{entities $x \in \mathcal{X}$}
            \State Add $x$ to $N_{\mathcal{X}}$
        \EndFor
        \ForAll{visits $\mathcal{S}_t^{(i)} \in \mathcal{R}$}
            \State $e^{\mathcal{X}} \gets$ hyperedge for entities in $\mathcal{S}_t^{(i)}$ of type $\mathcal{X}$
            \State Add $e^{\mathcal{X}}$ to $E_{\mathcal{X}}$
        \EndFor
    \EndProcedure
    \end{algorithmic}
    \end{algorithm}

    \subsection{EHR Hypergraph Construction}
    When constructing the EHR hypergraph, we do not merge all medical entities into a single heterogeneous hypergraph. Instead, Alg.~\ref{alg:hypergraph} outlines the process for building separate hypergraphs for each domain: diagnosis, procedure, and medication. This design is motivated by three factors: context specificity, complexity management, and scalability, as detailed below.
    \begin{itemize}
        \item \textbf{Context Specificity}: Diagnoses, procedures, and medications each carry distinct medical information and contextual meanings. By constructing separate hypergraphs, we can capture the unique contextual relationships inherent to each type of medical entity more precisely.
        \item \textbf{Complexity Management}: A single visit may contain as many as hundreds of medical entities. Incorporating all these entities into a single heterogeneous hyperedge could lead to excessively high complexity and introduce additional noise, making it more difficult to capture the relationships among nodes. Constructing separate hypergraphs reduces system complexity, making the model easier to train.
        \item \textbf{Flexibility and Scalability}: The structure of independently constructed hypergraphs offers greater flexibility for future model iterations. This design also facilitates extension to new entity types or updated datasets.
    \end{itemize}

    Taking the diagnosis hypergraph $\mathcal{H}_{\mathcal{D}} = (\mathcal{N}_\mathcal{D}, \mathcal{E}_\mathcal{D})$ as an example, it is constructed in two steps:
    \begin{itemize}
        \item \textbf{Node Set Construction}: Create nodes for all diagnostic entities in $\mathcal{D}$ and add them to the hypergraph node set $\mathcal{N}_\mathcal{D}$.
        \item \textbf{Hyperedge Set Construction}: Traverse all visits in the EHR training data. For each diagnostic vector $\mathbf{d}_t^{(i)}$ in visit $\mathcal{S}_t^{(i)}$, create a hyperedge $e$ connecting all diagnostic entities contained in that visit, and add $e$ to the hypergraph hyperedge set $\mathcal{E}_\mathcal{D}$.
    \end{itemize}
    The construction procedures for $\mathcal{H}_{\mathcal{P}}$ and $\mathcal{H}_{\mathcal{M}}$ follow the same logic.

    \section{Framework}
\begin{figure*}[htb]
        \centering
        \includegraphics[width=\textwidth, keepaspectratio]{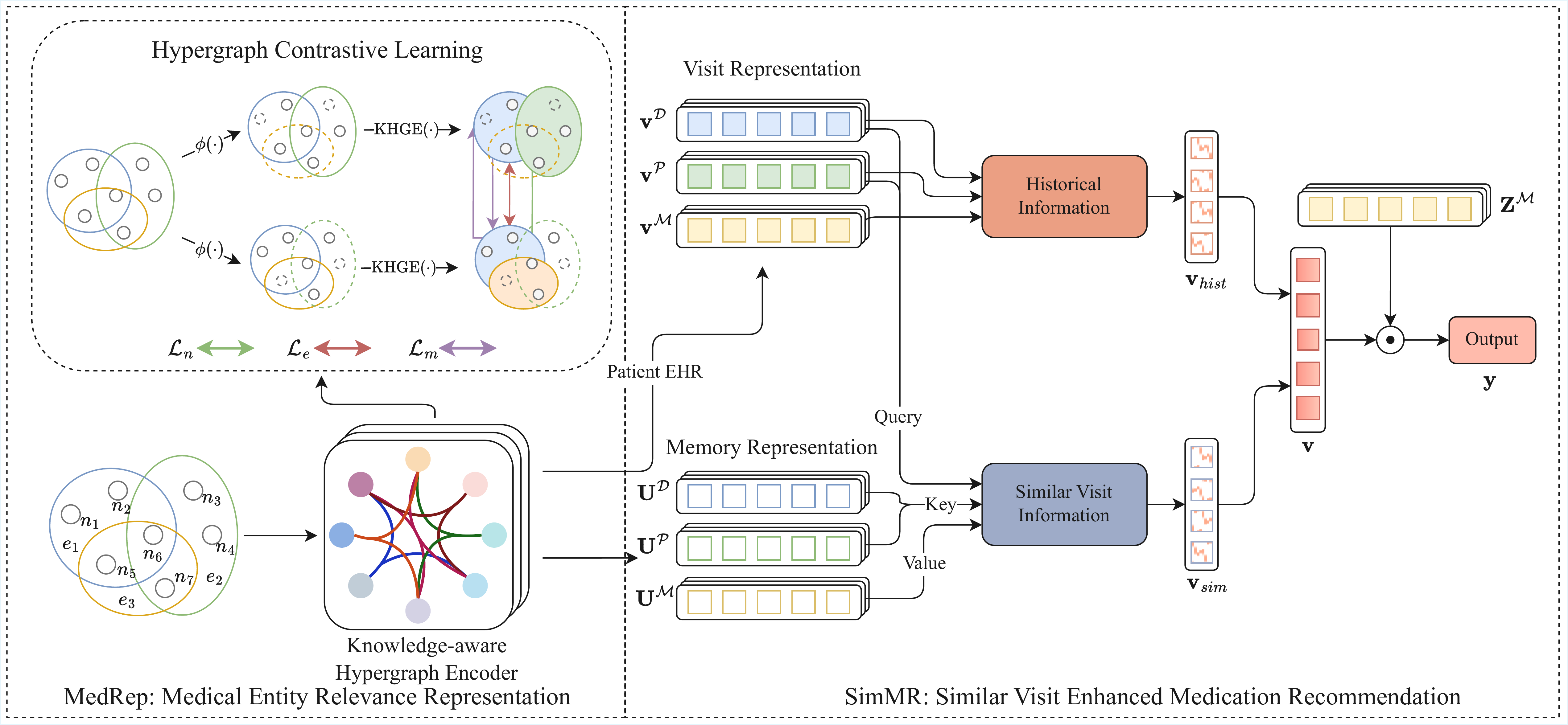}\\
        \caption{
            Overall architecture of \Model. \Model~comprises two stages: the Medical Entity Relevance Representation Stage (\MedRep) and the Similar Visit Enhanced Medication Recommendation Stage (\MedRec). 
            \rev{\MedRep~focuses on encoding intra-visit set-level combinatorial semantics into a globally consistent, retrieval-friendly embedding space. \MedRec~integrates longitudinal history and visit-conditioned retrieved similar visits to refine latent condition estimation.}
        }
        \label{fig:HypeMed}
    \end{figure*}

    This section outlines the \Model~framework, as illustrated in Fig.~\ref{fig:HypeMed}. The architecture primarily comprises the Medical Entity Relevance Representation Stage (\MedRep) and the Similar Visit Enhanced Medication Recommendation Stage (\MedRec). 
    We also explored an end-to-end variant in early experiments, which resulted in an approximately 3\% decrease in Jaccard score and lower training efficiency. Therefore, in this paper we adopt a two-stage training pipeline.

    We adopt a two-stage training paradigm, consisting of pre-training followed by fine-tuning. 
    First, we pre-train \MedRep\ on the clinical hypergraph with a contrastive objective to learn medical \emph{entity} (node) and \emph{visit} (hyperedge) representations.  
    {At the start of \MedRec, we instantiate two \emph{trainable} embedding tables for entities and visit hyperedges, initialized with the pre-trained \MedRep\ outputs. 
    During fine-tuning, these embedding tables and all recommender parameters are optimized end-to-end under the recommendation loss.}
    This preserves the semantic structure learned in pre-training while allowing task-specific adaptation at the embedding level, yielding stable optimization and improved downstream performance.
    The overall training pipeline is illustrated in Alg.~\ref{alg:training_pipeline}.

    \begin{algorithm}
    \caption{Two-Stage Training Pipeline for \Model}
    \label{alg:training_pipeline}
    \begin{algorithmic}[1]
    \Function{TrainModel}{$\mathcal{H}, \mathcal{D}, \mathcal{P}, \mathcal{M}$}
        \State Initialize KHGE parameters \Comment{Stage 1: \MedRep~training}
        \State Construct augmented subgraphs $\mathcal{H}_1, \mathcal{H}_2$ by dropout on $\mathcal{H}$
        \For{$epoch = 1$ to $e_{rep}$}
            \State $(\mathbf{Z}_1, \mathbf{U}_1) \gets \Call{EncodeHypergraph}{\mathcal{H}_1}$ 
                \Comment{{Encode augmented hypergraph via the hypergraph encoder (Sec.~\ref{sec:KHGE})}}
            \State $(\mathbf{Z}_2, \mathbf{U}_2) \gets \Call{EncodeHypergraph}{\mathcal{H}_2}$ 
                \Comment{{Encode augmented hypergraph via the hypergraph encoder (Sec.~\ref{sec:KHGE})}}
            \State Optimize $\mathbf{Z}_1,\mathbf{U}_1,\mathbf{Z}_2,\mathbf{U}_2$ using contrastive loss $\mathcal{L}_{cl}$ \Comment{{Eq.~(5)}}
        \EndFor
        \State \textbf{return} $\mathbf{Z}, \mathbf{U}$ \Comment{Final entity and hyperedge embeddings}

        \vspace{5mm}
        \State Initialize \MedRec~with $\mathbf{Z}, \mathbf{U}$  \Comment{{Stage 2: \MedRec~training (embeddings initialized from \MedRep~and trainable)}}
        \For{$epoch = 1$ to $e_{rec}$}
            \ForAll{visits $\mathcal{S}^{(i)}_t$}
                \State Retrieve $k$ similar visits for $\mathbf{v}_{\mathrm{sim}}$
                \State $\mathbf{v}_{\mathrm{hist}} \gets \Call{ComputeHistoricalRepresentation}{\mathcal{S}^{(i)}_t}$ 
                    \Comment{{Aggregate past visits via temporal attention (Sec.~\ref{sec:hist_rep})}}
                \State $\mathbf{v}_{\mathrm{sim}} \gets \Call{ComputeSimilarVisitRepresentation}{\mathcal{S}^{(i)}_t}$ 
                    \Comment{{Encode retrieved similar visits (Sec.~\ref{sec:sim_rep})}}
                \State Fuse $\mathbf{v}_{\mathrm{hist}}$ and $\mathbf{v}_{\mathrm{sim}}$ to obtain $\mathbf{v}_t$
                \State Optimize with total loss $\mathcal{L}$
            \EndFor
        \EndFor
        \State \textbf{return} trained \MedRec~model
    \EndFunction
    \end{algorithmic}
    \end{algorithm}

    \subsection{\MedRep: Medical Entity Relevance Representation}
    In the \MedRep~stage, we obtain node and hyperedge representations through hypergraph-based contrastive learning. We first generate two augmented subgraphs of the original hypergraph and then apply a knowledge-aware hypergraph encoder (KHGE) to compute node and hyperedge embeddings for both views. Finally, we perform contrastive learning between these two views to enhance embedding quality. \rev{By modeling each visit as a hyperedge that jointly connects diagnoses, procedures, and medications, \MedRep\ mitigates the intra-visit semantic fragmentation (Introduction) and yields context-aware entity and visit embeddings.} Below we first introduce the encoder, followed by the augmentation and contrastive objectives.

    \subsubsection{Knowledge-aware Hypergraph Encoder}
    \label{sec:KHGE}
\begin{figure}[htb]
        \centering
        \includegraphics[width=0.9\textwidth]{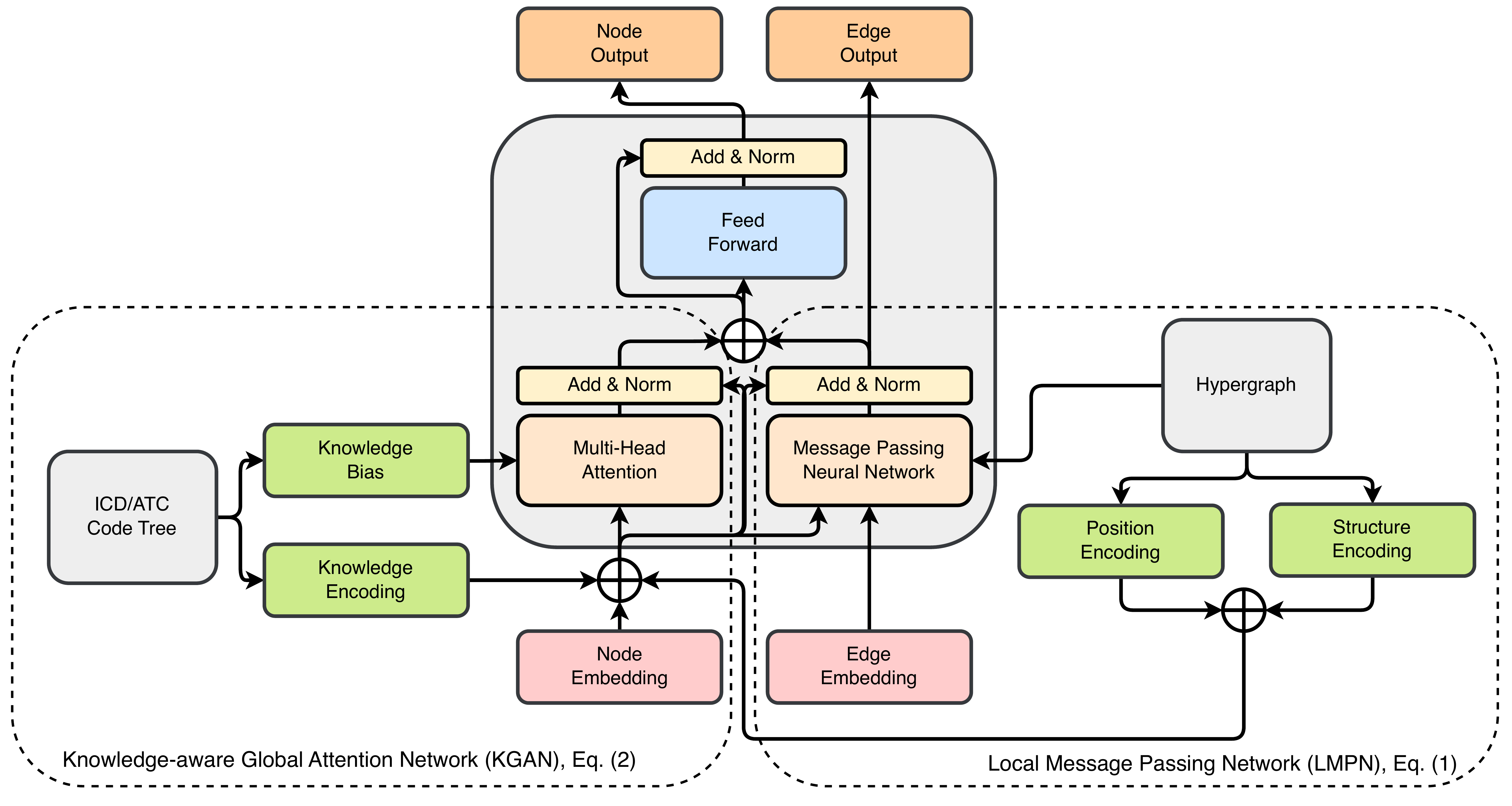}
        \caption{
            The detailed architecture of the Knowledge-aware Hypergraph Encoder (KHGE).
            The encoder consists of a Local Message Passing Network (LMPN) and a Knowledge-aware Global Attention Network (KGAN).
            The LMPN focuses on message propagation over the medical hypergraph, while the KGAN integrates medical knowledge into a global self-attention mechanism.
            The two outputs are combined through a feed-forward fusion layer with residual connections.
        }
        \label{fig:KHGE}
    \end{figure}

    We design a knowledge-aware hypergraph encoder (Fig.~\ref{fig:KHGE}) to capture contextual information and incorporate prior medical category knowledge. 
    This encoder consists of two complementary modules: (1) a \textit{Local Message Passing Network (LMPN)} that models neighborhood-level dependencies on the medical hypergraph, and 
    (2) a \textit{Knowledge-aware Global Attention Network (KGAN)} that incorporates structured medical knowledge (e.g., ICD/ATC hierarchies) to capture global correlations. 
    The outputs of both components are fused via a two-layer feed-forward network with residual connections and layer normalization. The detailed formulations for each module are presented as follows.

    \noindent \textbf{(a) Local Message Passing Network (LMPN).}
    Research on graph neural networks~\cite{GNN} and hypergraph neural networks~\cite{hgnn} suggests that node representations derived from local message passing effectively model neighborhood relationships, making nodes connected by edges more similar.
    Thus, applying local message passing on a medical hypergraph enhances the similarity of medical entity representations within the same visit. 
    The LMPN is implemented via a hypergraph attention layer followed by layer normalization and residual connections, as illustrated in Fig.~\ref{fig:KHGE}. 
    Local message passing involves two processes: hyperedge representation computation and node representation computation:

    \begin{equation}
        \begin{aligned}
            \mathbf{u}_j^{(k+1)} &= \sum_{n_i \in \mathcal{N}(e_j)} \alpha_{ij}^{(k)} \mathbf{z}_i^{(k)}, \\
            \mathbf{z}_i^{(k+1)} &= \sum_{e_j \in \mathcal{N}(n_i)} \alpha_{ij}^{(k)} \mathbf{u}_j^{(k)},
        \end{aligned}
        \label{eq:LMPN}
    \end{equation}
    where the attention coefficient $\alpha_{ij}^{(k)}$ controls the message weight between node $n_i$ and hyperedge $e_j$. Before aggregation, a learnable linear transformation is applied to refine both node and hyperedge embeddings.

    \noindent \textbf{(b) Knowledge-aware Global Attention Network (KGAN).}
    To capture long-range dependencies and inject domain knowledge, we introduce the Knowledge-aware Global Attention Network (KGAN).
    Medical knowledge from ICD or ATC code hierarchies provides a structured prior indicating semantic closeness among entities.
    We first transform the hierarchical tree into a knowledge bias matrix $\bm{\Omega}$, where each element encodes the path distance between two entities and is mapped to a learnable bias parameter. 
    This bias is then incorporated into the multi-head self-attention mechanism as:
    \begin{equation}
        \begin{aligned}
            \mathbf{Z}_g^{(k+1)} =
            \mathrm{Softmax}\!\left(
            \frac{\mathbf{W}_q \mathbf{Z}^{(k)} (\mathbf{W}_k \mathbf{Z}^{(k)})^{\mathrm{T}} + \bm{\Omega}}{\sqrt{d}}
            \right)
            \mathbf{W}_v \mathbf{Z}^{(k)}.
        \end{aligned}
        \label{eq:KGAN}
    \end{equation}
    Here, $\mathbf{W}_q$, $\mathbf{W}_k$, and $\mathbf{W}_v$ are learnable projection matrices, and $\bm{\Omega}$ provides a prior bias that allows the attention to favor medically related entities.

    \noindent \textbf{(c) Fusion and Multi-layer Aggregation.}
    We integrate the node representations from LMPN and KGAN through a two-layer feed-forward network (FFN) with residual connections and layer normalization, where $\mathbf{Z}_\ell^{(k + 1)}$ and $\mathbf{Z}_g^{(k + 1)}$ denote the outputs of LMPN and KGAN, respectively. To mitigate over-smoothing, we perform a layer-wise average of intermediate representations:
    \begin{equation}
        \begin{aligned}
            &\mathbf{Z}^{(k+1)} = \mathrm{FFN}\!\left(\mathbf{Z}_\ell^{(k+1)} + \mathbf{Z}_g^{(k+1)}\right), \\
            &\mathbf{Z} = \tfrac{1}{L}\sum_{k=1}^{L}\mathbf{Z}^{(k)}, \quad
            \mathbf{U} = \tfrac{1}{L}\sum_{k=1}^{L}\mathbf{U}^{(k)}.
        \end{aligned}
        \label{eq:KHGE_2}
    \end{equation}
    This fusion mechanism effectively combines local relational learning with global knowledge reasoning. 

    \subsubsection{Hypergraph Contrastive Learning}
    Existing research~\cite{TriCL} indicates that hypergraph contrastive learning can enhance the representational capabilities of embeddings. Hypergraph contrastive learning brings the representations of similar nodes and hyperedges closer together while distancing those of dissimilar nodes and hyperedges. This aligns with our expectations for medical entity representation. Hence, we employ a hypergraph contrastive learning approach based on nodes and hyperedges for the training phase of \MedRep. Initially, we construct two augmented subgraphs based on the original hypergraph. Subsequently, we compute node and hyperedge representations based on these two subgraphs. Finally, contrastive learning is executed between these two sets of representations.

    \noindent \textbf{Augmented Subgraphs Construction}.
    EHR data often contains a considerable amount of noise, which can negatively impact the learning of medical entity representations. To address this issue, constructing augmented subgraphs through random dropout of elements in the original visits has become crucial for mitigating the effects of noise~\cite{TriCL, SGL}.
    We independently perform two stochastic augmentations on the original hypergraph to obtain subgraphs $\mathcal{H}_1$ and $\mathcal{H}_2$, by randomly dropping nodes, incidences, and features with predefined probabilities. Dropped elements are excluded when computing the contrastive objectives.

    \noindent \textbf{Contrastive Learning.}
    After obtaining two augmented subgraphs $\mathcal{H}_1$ and $\mathcal{H}_2$, 
    we use the knowledge-aware hypergraph encoder to generate node and hyperedge embeddings for both views:
    \begin{equation}
        \begin{aligned}
            \mathbf{Z}_1, \mathbf{U}_1 &= \mathrm{KHGE}(\mathcal{H}_1), \\
            \mathbf{Z}_2, \mathbf{U}_2 &= \mathrm{KHGE}(\mathcal{H}_2).
        \end{aligned}
        \label{eq:HCL_1}
    \end{equation}

    To enhance the consistency of multi-level representations, 
    we adopt a three-level contrastive learning strategy encompassing node-, hyperedge-, and membership-level objectives 
    (as illustrated in the upper-left of Fig.~\ref{fig:HypeMed}). 
    This design ensures that (1) nodes sharing similar contexts are mapped closely in the embedding space, 
    (2) hyperedges containing related nodes exhibit similar representations, and 
    (3) membership contrast bridges the alignment between node and hyperedge embeddings. 
    The overall contrastive objective is formulated as:
    \begin{equation}
        \mathcal{L}_{cl} =
        \mathrm{InfoNCE}(\mathbf{Z}_1, \mathbf{Z}_2)
        + \lambda_1\,\mathrm{InfoNCE}(\mathbf{U}_1, \mathbf{U}_2)
        + \lambda_2\,\mathrm{InfoNCE}(\mathbf{Z}_1, \mathbf{U}_2),
        \label{eq:HCL_2}
    \end{equation}
    where $\lambda_1$ and $\lambda_2$ balance the contributions of hyperedge-level and membership-level contrastive terms.
    The InfoNCE loss~\cite{InfoNCE} is defined as:
    \begin{equation}
        \mathrm{InfoNCE}(\mathbf{U}, \mathbf{V})
        = -\tfrac{1}{N}\sum_{i=1}^{N}
        \log\!\frac{\exp(\mathbf{u}_i\!\cdot\!\mathbf{v}_i/\tau)}
        {\sum_{j=1}^{N}\exp(\mathbf{u}_i\!\cdot\!\mathbf{v}_j/\tau)},
        \label{eq:HCL_3}
    \end{equation}
    where $\tau$ is the temperature parameter controlling the smoothness of the contrastive distribution.

    During the \MedRep~stage, we optimize $\mathcal{L}_{cl}$ to jointly learn 
    node embeddings $\mathbf{Z}$ (medical entities) and hyperedge embeddings $\mathbf{U}$ (visits), 
    thereby establishing a unified and semantically consistent representation space.

    {Upon completing \MedRep, we export the learned embedding matrices for medical entities and visit embeddings (factorized into diagnosis, procedure, and medication subspaces). 
    These matrices are then used in \MedRec\ to \emph{initialize} two \emph{trainable} embedding tables (entities and visits), 
    which are updated jointly with the recommender parameters. Specifically, the task-specific representations are obtained as:}
    \begin{equation}
        \mathbf{Z}_{\mathcal{X}}, \mathbf{U}_{\mathcal{X}}
        = \mathrm{KHGE}(\mathcal{H}_{\mathcal{X}}),
        \quad \mathcal{X} \in \{\mathcal{D}, \mathcal{P}, \mathcal{M}\}.
        \label{eq:HCL_4}
    \end{equation}
    In this way, the pre-trained embeddings serve as a high-quality initialization for the recommendation task, enabling the model to leverage rich medical correlations captured during the contrastive learning phase.

    \subsection{\MedRec: Similar Visit Enhanced Medication Recommendation}
    In the second stage, we train \MedRec~based on the representations obtained from \MedRep, specifically optimizing for the medication recommendation task. In clinical practice, similar visits are often consulted as external evidence~\cite{brooks1991role,solomon2006case,brown2016patient}. Accordingly, \rev{\MedRec~refines latent condition estimation for the current visit by augmenting it with two types of inter-visit references: (i) information from the patient trajectory (temporal causality) and (ii) visit-conditioned retrieved similar visits (analogical evidence).}

    We first compute the representation of the current visit and then feed it into two channels to obtain complementary visit representations: a \emph{historical} channel and a \emph{similar-visit} channel. Finally, we fuse these two representations and apply a dot-product scorer to produce the medication probabilities. The pipeline is shown in the right panel of Fig.~\ref{fig:HypeMed}.

    \subsubsection{Visit Representation}

    A single visit may involve multiple domains of medical entities (diagnoses, procedures, medications) that exhibit semantic gaps. We therefore compute domain-specific visit representations for each visit.

    We first mean-pool the entity embeddings in a given domain to form an initial visit representation ($\mathbf{z}_t^\mathcal{X}$), then refine it via attention between it and the corresponding entity embeddings (like hyperedge computation in \MedRep). This yields the domain-specific visit representation and facilitates subsequent similar-visit retrieval.

    Concretely, for domain $\mathcal{X}\!\in\!\{\mathcal{D},\mathcal{P},\mathcal{M}\}$, we compute
    \begin{equation}
        \mathbf{v}_t^\mathcal{X}
        = \mathrm{MHA}\!\left(\mathbf{z}_t^\mathcal{X}, \mathbf{Z}^\mathcal{X}, \mathbf{Z}^\mathcal{X}\right),
        \label{eq:visit_rep_equation}
    \end{equation}
    where $\mathrm{MHA}(\cdot)$ denotes multi-head attention and $\mathbf{z}_t^\mathcal{X}$ is the mean-pooled embedding of the entities in domain $\mathcal{X}$ for the $t$-th visit. We obtain $\mathbf{v}_t^{\mathcal{D}}$, $\mathbf{v}_t^{\mathcal{P}}$, and $\mathbf{v}_t^{\mathcal{M}}$ accordingly, and denote the medication label as $\mathbf{m}_t \in \{0,1\}^{|\mathcal{M}|}$.

    \subsubsection{Historical Information Representation}
    \label{sec:hist_rep}
    The success of longitudinal medication recommendation methods~\cite{GAMENet, SafeDrug, COGNet} underscores the significance of historical information in the task of medication recommendation. This subsection describes how we compute visit representations using historical information.

    Two types of information are particularly crucial in a patient trajectory. The first is the patient's health status information, which includes diagnostic and procedure conditions. These visits directly reflect the patient's health condition. The second type is medication information. Existing studies~\cite{MICRON, COGNet} indicate that patients exhibit continuity in medication usage. Medications previously administered are likely to be used again in the future. 
    Hence, modeling medication information can reflect the patient's historical trends in medication usage.

    However, considering all historical visits in the patient trajectory is impractical, as not all historical visits are relevant to the current one. Visits from the distant past may not provide helpful references for current medication. Therefore, we employ a window-based visit attention mechanism over the patient trajectory. The following formula can represent this process:

    \begin{equation}
        \mathbf{v}_{\mathrm{hist}} =
        \mathrm{MHA}\!\left(\mathbf{h}_t, \mathbf{V}_{t-k:t-1}, \mathbf{V}_{t-k:t-1}\right),
        \label{eq:medrec_hist}
    \end{equation}
    where $\mathbf{h}_t$ is the current health-status embedding (from diagnoses and procedures), and $\mathbf{V}_{t-k:t-1}\!\in\!\mathbb{R}^{k\times d}$ stacks the $k$ most recent visit embeddings. The resulting $\mathbf{v}_{\mathrm{hist}}$ captures recent diagnostic, procedural, and medication trends within the temporal window.

    \subsubsection{Similar Visit Information Representation}
    \label{sec:sim_rep}
    In the \MedRep\ stage, we obtain embeddings for all training visits (hyperedges) and factorize them into domain-specific subspaces. Given $\mathbf{h}_t$, we retrieve the top-$k$ similar visits in the health-status subspace to ensure consistency with the pre-trained geometry. 
    \rev{This design is clinically motivated: we retrieve similar past visits as external evidence, and condition the retrieval on the current visit's health-status embedding $\mathbf{h}_t$ to ensure relevance.}
    Let $\mathbf{U}^{H}$ and $\mathbf{U}^{\mathcal{M}}$ denote the hyperedge embeddings in the health-status (diagnosis/procedure) and medication subspaces, respectively. We aggregate the retrieved evidence via attention:
    \begin{equation}
        \mathbf{v}_{\mathrm{sim}} =
        \mathrm{MHA}\!\left(\mathbf{h}_t, \mathbf{U}^{H}_{\text{top-}k}, \mathbf{U}^{\mathcal{M}}_{\text{top-}k}\right),
        \label{eq:sim_case}
    \end{equation}
    where $\mathbf{U}^{H}_{\text{top-}k}$ and $\mathbf{U}^{\mathcal{M}}_{\text{top-}k}$ are from the retrieved $k$ nearest visits in the corresponding subspaces.

    To further improve retrieval alignment, we adopt an in-batch contrastive objective that aligns the current visit's embeddings with those of the retrieved visits in the same representation spaces:
    \begin{equation}
        \mathcal{L}_{cl}^{sim} =
        \mathrm{InfoNCE}\!\left(\mathbf{h}_t, \mathbf{U}^{H}\right)
        + \mathrm{InfoNCE}\!\left(\mathbf{v}_t^{\mathcal{M}}, \mathbf{U}^{\mathcal{M}}\right),
        \label{eq:train_cl}
    \end{equation}
    \rev{where both terms use temperature-scaled inner-product similarity, consistent with the pre-trained geometry, and help mitigate representation–retrieval mismatch during fine-tuning by maintaining geometric consistency between the current visit and the retrieved visits.}

    \subsubsection{Channels Fusion} 
    We perform representation fusion after obtaining the visit representations from the two channels (historical information and similar visit information). We stack the two channel representations for the same visit and compute their weights using a two-layer MLP followed by a Softmax layer. This process is formulated as follows:
    \begin{equation}
        \mathbf{v}_t
        = \alpha_{\mathrm{hist}}\,\mathbf{v}_{\mathrm{hist}}
        + \alpha_{\mathrm{sim}}\,\mathbf{v}_{\mathrm{sim}},
        \quad
        [\alpha_{\mathrm{hist}},\alpha_{\mathrm{sim}}]
        = \mathrm{Softmax}\!\big(\mathrm{MLP}([\mathbf{v}_{\mathrm{hist}} \| \mathbf{v}_{\mathrm{sim}}])\big).
        \label{eq:channel_fusion_equation_1}
    \end{equation}
    To capture information from different perspectives, we apply regularization constraints to them. We encourage the two representations to be decorrelated (approximately orthogonal). Concretely, we minimize their cosine similarity:
    \begin{equation}
        \mathcal{L}_{orth} =
        \left|\frac{\mathbf{v}_{\mathrm{hist}}\!\cdot\!\mathbf{v}_{\mathrm{sim}}}
        {\|\mathbf{v}_{\mathrm{hist}}\|\;\|\mathbf{v}_{\mathrm{sim}}\|}\right|.
        \label{eq:channel_fusion_equation_2}
    \end{equation}

    \subsubsection{Prediction and Objectives}
    We calculate the probability distribution for recommending medications by assessing the dot-product similarity between the representations of patient visits and medications. This can be expressed as follows:
    \begin{equation}
        \mathbf{y}_t = \sigma\!\left(\mathbf{v}_t^{\top}\mathbf{Z}^{\mathcal{M}}\right),
        \label{predict_equation_1}
    \end{equation}
    where $\sigma(\cdot)$ is the logistic function applied element-wise. Given a threshold $\eta$, the predicted medication set is
    \begin{equation}
        \hat{m}_t = \{\, i \mid (\mathbf{y}_t)_i \ge \eta \,\}.
    \end{equation}
    Here, $({\mathbf{y}_t})_i$ denotes the $i^{th}$ element of ${\mathbf{y}_t}$. That is, $\hat{m}_t$ is the set of indices whose predicted probabilities exceed $\eta$.

    We adopt the binary cross-entropy loss, multi-label loss, and DDI loss used in SafeDrug~\cite{SafeDrug}, where the binary cross-entropy loss and multi-label loss are related to prediction accuracy, and the DDI loss pertains to the DDI rate. These loss functions are computed as follows:
    \begin{itemize}
        \item \textbf{Binary Cross Entropy Loss} is a typical loss function used in machine learning, particularly effective in multi-label classification settings. It calculates the loss by measuring the distance between the model’s predicted probabilities and the actual binary labels, penalizing predictions based on the divergence from the true labels. This makes it highly suitable for scenarios where each instance can belong to multiple classes independently.
        \begin{equation}
            \mathcal{L}_{bce}
            = -\sum_{i=1}^{|\mathcal{M}|}\Big[\mathbf{m}_{t,i}\log\mathbf{y}_{t,i}
            +(1-\mathbf{m}_{t,i})\log(1-\mathbf{y}_{t,i})\Big].
            \label{sub_loss_1}
        \end{equation}
        \item \textbf{Multi-label Loss} promotes model predictions where the presence of one label significantly exceeds the probability of an absent label by a specified margin, effectively learning both label occurrence and their relationships in multi-label classification tasks.
        \begin{equation}
            \mathcal{L}_{multi}
            = \tfrac{1}{|\mathcal{M}|}
            \!\!\sum_{i,j:\mathbf{m}_{t,i}=1,\mathbf{m}_{t,j}=0}
            \!\!\max\!\left(0,\,1-(\mathbf{y}_{t,i}-\mathbf{y}_{t,j})\right).
            \label{sub_loss_2}
        \end{equation}
        \item \textbf{DDI Loss} is designed to minimize the risk of adverse drug-drug interactions in predictive models. It penalizes predicted probabilities assigned to medication pairs with known DDIs, thus reducing the likelihood of recommending harmful medication combinations in clinical settings.
        \begin{equation}
            \mathcal{L}_{ddi}
            = \sum_{i,j}\mathbf{A}_{ij}\,\mathbf{y}_{t,i}\,\mathbf{y}_{t,j}.
            \label{sub_loss_3}
        \end{equation}
    \end{itemize}
    Here, $|\mathcal{M}|$ is the number of medications, $\mathbf{m}_t$ is the ground-truth multi-hot label, $\mathbf{y}_t$ is the predicted probability vector, and $\mathbf{A}\!\in\!\mathbb{R}^{|\mathcal{M}|\times|\mathcal{M}|}$ is the DDI adjacency matrix.
    The total loss function can be formulated as follows:
    \begin{equation}
        \mathcal{L} =
        \mathcal{L}_{bce}
        + \lambda_{multi}\,\mathcal{L}_{multi}
        + \lambda_{ddi}\,\mathcal{L}_{ddi}
        + \lambda_{aux}\,(\mathcal{L}_{cl}^{sim} + \mathcal{L}_{orth}),
        \label{total_loss_equation}
    \end{equation}
    where $\lambda_{multi}$, $\lambda_{ddi}$, and $\lambda_{aux}$ are balancing hyperparameters.

    \section{Experiments}
    To evaluate the performance of \Model~against state-of-the-art (SOTA) models, we conduct extensive experiments on three widely used and publicly available benchmark datasets: MIMIC-III~\cite{MIMIC-III}, MIMIC-IV~\cite{MIMIC-IV}, and {eICU~\cite{eicu}}.

    \subsection{Experimental Settings}
    \subsubsection{Datasets.} 
    MIMIC-III\footnote{\href{https://physionet.org/content/mimiciii/1.4/}{https://physionet.org/content/mimiciii/1.4/}} and MIMIC-IV\footnote{\href{https://physionet.org/content/mimiciv/2.0/}{https://physionet.org/content/mimiciv/2.0/}} are large-scale de-identified EHR databases from the Beth Israel Deaconess Medical Center, comprising data from over 40{,}000 ICU patients across diverse clinical conditions. 
    {\textbf{eICU}}\footnote{\href{https://physionet.org/content/eicu-crd/2.0/}{https://physionet.org/content/eicu-crd/2.0/}} {is a multi-center critical care database covering more than 200 hospitals across the United States and is used to assess model generalizability under cross-institutional settings.}

    Following prior work~\cite{COGNet, Carmen, DGCL, yang2023molerec}, we adopt the data preprocessing pipeline from SafeDrug~\cite{SafeDrug}, which includes filtering low-frequency entities, removing patients with fewer than two visits, and converting codes into standardized ICD and ATC formats. 
    {For eICU, medications are recorded using GTC codes. Since procedure codes are unavailable, we use only diagnosis and medication information for this dataset. Moreover, because eICU lacks a DDI graph, we exclude DDI-related losses and metrics during both training and evaluation.}

    The statistics of the processed datasets are summarized in Tab.~\ref{tab:datasets}. 
    {Following common practice in contrastive representation learning (e.g., TriCL~\cite{TriCL} and SGL~\cite{SGL}), we construct two independently augmented views of each hypergraph to form positive pairs for contrastive supervision. Empirically, this setup achieves an optimal balance between representation diversity and computational efficiency.}
    \begin{table}[htb]
        \centering
        \caption{Statistics of processed MIMIC-III/MIMIC-IV and {eICU}.}
        \begin{tabularx}{0.9\textwidth}{>{\centering\arraybackslash}X>{\centering\arraybackslash}X>{\centering\arraybackslash}X>{\centering\arraybackslash}X}
        \toprule
        \textbf{Item} & \textbf{MIMIC-III} & \textbf{MIMIC-IV} & \textbf{eICU}\\
        \midrule
        
        \# of patients & 6,350 & 9,036 & 7,855\\
            \# of visits & 15,032 & 20,616 & 16,869\\
        avg. \# of visits & 2.37 & 2.28 & 2.15\\
        
        \# of unique diag. codes & 1,958 & 1,892 & 692\\
        \# of unique proc. codes & 1,430 & 4,939 & -\\
        \# of unique med. codes & 131 & 131 & 46\\

        avg. \# of diag. per visit & 10.51 & 13.62 & 4.32\\
        avg. \# of proc. per visit & 3.84 & 3.55 & -\\
        avg. \# of med. per visit & 11.44 & 10.29 & 10.59\\
        \bottomrule
        \end{tabularx}
        \label{tab:datasets}
    \end{table}

    \subsubsection{Evaluation.}
    To comprehensively evaluate model performance, we employ five standard metrics: Jaccard Similarity $\uparrow$ (\%), F1-score $\uparrow$ (\%), and Precision-Recall AUC (PRAUC) $\uparrow$ (\%) for accuracy; DDI Rate $\downarrow$ (\%) for safety; and the average number of prescribed medications (\# Med.) $\downarrow$ for prescription compactness. 
    The ground-truth DDI rates are 8.68\% and 7.24\% for the MIMIC-III and MIMIC-IV datasets, respectively.\footnote{Due to discrepancies across prior reports~\cite{COGNet, DGCL, DrugRec}, we recalculated the DDI ground truth following the standard procedure and provide reference code for reproducibility.} 
    Given that the DDI ground truth is non-zero, our objective is not to eliminate DDIs entirely, but to ensure that the model’s recommended medication combinations exhibit a DDI rate below the empirical human benchmark. Symbols $\uparrow$ and $\downarrow$ respectively indicate \textit{higher is better} and \textit{lower is better}.
    To ensure a fair and consistent comparison, we adopt the conventional evaluation metrics used in the domain of medication recommendation~\cite{GAMENet, SafeDrug, COGNet}. The calculation methods for each metric are as follows:
    \begin{itemize}
        \item \textbf{Jaccard Similarity} between the predicted and ground-truth medication sets for an individual patient is computed as:
        \begin{equation}
            \mathrm{Jaccard} = 
            \frac{1}{T}\sum_{t=1}^{T}
            \frac{|m_t \cap \hat{m}_t|}{|m_t \cup \hat{m}_t|},
            \label{eq:jaccard}
        \end{equation}
        where $m_t$ denotes the ground-truth medication set and $\hat{m}_t$ the predicted set at visit $t$.
        
        \item \textbf{F1-score} is defined as:
        \begin{equation}
            \mathrm{F1} = 
            \frac{1}{T}\sum_{t=1}^{T}
            \frac{2 P_t R_t}{P_t + R_t},
            \quad
            P_t = \frac{|m_t \cap \hat{m}_t|}{|\hat{m}_t|}, \quad
            R_t = \frac{|m_t \cap \hat{m}_t|}{|m_t|}.
            \label{eq:f1}
        \end{equation}
        Here, $P_t$ and $R_t$ denote the precision and recall for visit $t$, respectively, and $T$ is the total number of visits.
        
        \item \textbf{PRAUC} represents the area under the precision–recall curve. Following prior work~\cite{COGNet, SafeDrug}, we compute it as the mean of per-visit average precision scores:
        \begin{equation}
            \mathrm{PRAUC} = 
            \frac{1}{T}\sum_{t=1}^{T}\mathrm{AP}_t,
            \label{eq:prauc}
        \end{equation}
        where $\mathrm{AP}_t$ is the average precision of visit $t$.
        
        \item \textbf{DDI Rate} measures the proportion of predicted medication pairs that have known drug–drug interactions:
        \begin{equation}
            \mathrm{DDI} =
            \frac{1}{T}\sum_{t=1}^{T}
            \frac{\sum_{i=1}^{|\hat{m}_t|}\sum_{j=i+1}^{|\hat{m}_t|}\mathbb{1}\{\mathbf{A}_{\hat{m}_t^{(i)}, \hat{m}_t^{(j)}} = 1\}}
                {\sum_{i=1}^{|\hat{m}_t|}\sum_{j=i+1}^{|\hat{m}_t|}1},
            \label{eq:ddi}
        \end{equation}
        where $\mathbb{1}\{\cdot\}$ denotes the indicator function and $\mathbf{A}$ is the adjacency matrix of the DDI graph.
        
        \item \textbf{\# Med.} quantifies the average number of medications prescribed per visit:
        \begin{equation}
            \mathrm{\# Med.} = 
            \frac{1}{T}\sum_{t=1}^{T}|\hat{m}_t|.
            \label{eq:med_count}
        \end{equation}
    \end{itemize}

    First, we calculate these metrics for each patient, and then we compute the average value across all patients to determine the final performance of the method.

    Following prior work~\cite{SafeDrug, COGNet, Carmen, yang2023molerec}, we evaluate all models using a bootstrapping protocol rather than cross-validation. Each model is trained on a fixed training set with hyperparameters tuned on a validation set. During evaluation, we perform ten rounds of bootstrapped testing by randomly sampling 80\% of the test set with replacement, reporting the mean and standard deviation across rounds.

    \subsubsection{Baselines.}
    To demonstrate the effectiveness of \Model, we compare it against representative medication recommendation methods, grouped into three categories: \textit{machine-learning}, \textit{instance-based}, and \textit{longitudinal} approaches.

    \begin{itemize}
        \item \textbf{Machine Learning Methods.}
        \begin{itemize}
            \item \textbf{LR} is a logistic regression–based baseline that performs independent binary classification for each medication label.
            \item \textbf{ECC}~\cite{ECC} (Ensembled Classifier Chain) encodes diagnosis and procedure sets into multi-hot vectors and applies a chain of SVM classifiers to perform multi-label prediction. The chaining mechanism enables ECC to capture label dependencies and improve overall prediction accuracy.
        \end{itemize}

        \item \textbf{Instance-based Methods.}
        \begin{itemize}
            \item \textbf{LEAP}~\cite{LEAP} formulates medication recommendation as a sequential decision-making process. It employs a recurrent decoder with content-based attention to model label dependencies across visits.
        \end{itemize}

        \item \textbf{Longitudinal Methods.}
        \begin{itemize}
            \item \textbf{RETAIN}~\cite{RETAIN} is an interpretable predictive framework that processes EHR data using a reverse-time attention mechanism, enhancing both prediction accuracy and clinical interpretability.
            \item \textbf{GAMENet}~\cite{GAMENet} integrates a DDI knowledge graph with a memory-augmented GCN to jointly model patient history and medication safety.
            \item \textbf{SafeDrug}~\cite{SafeDrug} recommends effective and safe medication combinations by jointly modeling global and local molecular structures of medications. It also incorporates a controllable loss term to explicitly penalize DDIs.
            \item \textbf{COGNet}~\cite{COGNet} introduces a ``copy-or-predict'' mechanism within an encoder–decoder framework. By leveraging historical medication usage, COGNet determines whether to replicate past prescriptions or predict new ones.
            \item \textbf{Carmen}~\cite{Carmen} is a context-aware medication recommendation framework based on graph neural networks. It fuses molecular structure, contextual information, and DDI graph encoding to enhance both performance and safety.\footnote{We were unable to successfully execute Carmen's official implementation; therefore, we report the results provided in their paper~\cite{Carmen}, obtained under the same data preprocessing pipeline.}
            \item \textbf{DGCL}~\cite{DGCL} employs graph contrastive learning to jointly train medication knowledge graphs and EHR graphs, effectively controlling DDI levels and improving recommendation robustness.
            \item \textbf{VITA}~\cite{VITA} improves accuracy through relevant-visit selection and target-aware attention mechanisms, which better capture relationships between current and historical visits.
            \item \textbf{MoleRec}~\cite{yang2023molerec} utilizes molecular substructure–aware encoding and attention mechanisms to generate personalized and safer medication combinations.
            \item {\textbf{DAPSNet}~\cite{DAPSNet} employs dual attention mechanisms at both code and visit levels to construct comprehensive patient representations. It retrieves information from similar trajectories through a patient memory module and applies an information bottleneck to enhance robustness and safety.}

            \item {\textbf{PROMISE}~\cite{promise} introduces a pre-trained multimodal framework that integrates structured EHRs and clinical text via hypergraph and language encoders. By combining multimodal pre-training with controllable DDI optimization, it effectively balances accuracy and medication safety.}
        \end{itemize}
    \end{itemize}

    We adopted the hyperparameter configurations recommended in the original papers of all baseline models to ensure fair comparison. 
    For baselines without publicly available implementations, we carefully re-implemented them according to the methodological details described in their papers.

    Our method was implemented in Python~3.8.16 using PyTorch~1.13.1. 
    Experiments were conducted on a workstation equipped with dual Intel Xeon Gold~5318Y CPUs, 251GB of RAM, and three NVIDIA A40 GPUs. 
    The hyperparameter search space is summarized in Tab.~\ref{tab:setting}.
    \begin{table}[htb]
        \centering
        \caption{Hyperparameter search space.}
        \begin{tabularx}{0.75\textwidth}{>{\centering\arraybackslash}X>{\centering\arraybackslash}X}
        \toprule
        \textbf{Hyperparameter} & \textbf{Range} \\
        \midrule
        Learning rate & $[5\times10^{-3},\, 1\times10^{-4}]$ \\
        Weight decay & $[1\times10^{-4},\, 1\times10^{-5}]$ \\
        Dropout ratio & $[0,\, 0.5]$ \\
        Epochs & $50$ \\
        Batch size & $16$ \\
        Pretraining epochs & $\{150,\, 300,\, 500,\, 1500,\, 3000\}$ \\
        Pretraining learning rate & $\{1\times10^{-3},\, 5\times10^{-4}\}$ \\
        Pretraining weight decay & $\{1\times10^{-4},\, 1\times10^{-5}\}$ \\
        KHGE layers & $\{1,\, 2,\, 4\}$ \\
        Embedding dimension & $64$ \\
        $\lambda_{\text{multi}}$ & $[1,\, 10^{-6}]$ \\
        $\lambda_{\text{ddi}}$ & $[1,\, 10^{-6}]$ \\
        $\lambda_{\text{aux}}$ & $[1,\, 10^{-6}]$ \\
        \bottomrule
        \end{tabularx}
        \label{tab:setting}
    \end{table}

    \begin{table*}[htb]
        \centering
        \caption{Performance comparison on \textbf{MIMIC-III} in terms of Jaccard (\%), F1 (\%), PRAUC (\%), DDI (\%) and \# Med.. 
        Numbers in \textbf{bold} and \underline{underlined} indicate the best and the second-best performance, respectively, according to t-tests at the 95\% confidence level.
        The ground-truth rate of DDI in the MIMIC-III dataset is 8.68\%. 
        In the DDI column, values in {\color{orange!50!red}{orange}} indicate that the mean+std exceeds the ground truth (inferior), while values in {\color{blue!60!green}{blue}} indicate that the mean-std falls below it (superior).}
        \begin{tabularx}{\textwidth}{>{\centering\arraybackslash}X|
                                    >{\centering\arraybackslash}X
                                    >{\centering\arraybackslash}X
                                    >{\centering\arraybackslash}X
                                    >{\centering\arraybackslash}X
                                    >{\centering\arraybackslash}X}
            \Xhline{1pt}
            \textbf{Model} & \textbf{Jaccard}$\uparrow$ & \textbf{F1}$\uparrow$ & \textbf{PRAUC}$\uparrow$ & \textbf{DDI}$\downarrow$ & \textbf{\# Med.}$\downarrow$ \\
            \Xhline{0.5pt}
            LR              & $49.24_{\pm0.27}$   & $64.96_{\pm0.25}$   & $75.48_{\pm0.31}$   & $\color{blue!60!green}{8.30_{\pm0.07}}$       & $16.05_{\pm0.14}$ \\
            ECC             & $48.89_{\pm0.26}$   & $64.47_{\pm0.23}$   & $76.19_{\pm0.20}$   & $\color{blue!60!green}{8.57_{\pm0.08}}$       & $15.74_{\pm0.07}$ \\
            LEAP            & $45.50_{\pm0.18}$   & $61.68_{\pm0.17}$   & $65.46_{\pm0.42}$   & $\color{blue!60!green}{7.82_{\pm0.08}}$       & $18.57_{\pm0.08}$ \\
            \Xhline{0.2pt}
            RETAIN          & $48.60_{\pm0.34}$   & $64.66_{\pm0.34}$   & $75.98_{\pm0.43}$   & $\color{orange!50!red}{8.83_{\pm0.10}}$       & $18.92_{\pm0.13}$ \\
            GAMENet         & $50.18_{\pm0.20}$   & $65.84_{\pm0.19}$   & $76.48_{\pm0.23}$   & $\color{orange!50!red}{8.88_{\pm0.09}}$       & $27.39_{\pm0.19}$ \\
            SafeDrug        & $51.14_{\pm0.25}$   & $66.81_{\pm0.23}$   & $76.43_{\pm0.23}$   & $\color{blue!60!green}{5.87_{\pm0.04}}$       & $18.94_{\pm0.13}$ \\
            DAPSNet         & $51.68_{\pm0.30}$   & $67.22_{\pm0.26}$   & $76.64_{\pm0.40}$   & $\color{blue!60!green}{5.96_{\pm0.05}}$       & $21.14_{\pm0.15}$ \\
            PROMISE         & $52.33_{\pm0.25}$   & $67.91_{\pm0.21}$   & $\mathbf{80.58}_{\pm0.17}$   & $\color{orange!50!red}{9.33_{\pm0.15}}$       & $15.27_{\pm0.13}$ \\
            COGNet          & $52.94_{\pm0.39}$   & $68.30_{\pm0.36}$   & $76.79_{\pm0.25}$   & $\color{orange!50!red}{8.62_{\pm0.09}}$       & $28.08_{\pm0.18}$ \\
            Carmen          & $52.67_{\pm0.21}$   & $68.12_{\pm0.19}$   & $76.52_{\pm0.36}$   & -- & -- \\
            DGCL            & $52.54_{\pm0.15}$   & $67.98_{\pm0.14}$   & $77.28_{\pm0.15}$   & $\color{blue!60!green}{8.31_{\pm0.03}}$       & $21.92_{\pm0.05}$ \\
            VITA            & $52.15_{\pm0.25}$   & $67.42_{\pm0.21}$   & $76.23_{\pm0.25}$   & $\color{blue!60!green}{8.11_{\pm0.09}}$       & $29.53_{\pm0.15}$ \\
            MoleRec         & $\underline{53.12}_{\pm0.39}$ & $\underline{68.50}_{\pm0.33}$ & ${77.51}_{\pm0.25}$ & $\color{blue!60!green}{7.16_{\pm0.07}}$ & $20.54_{\pm0.16}$ \\
            \Xhline{0.2pt}
            \Model          & $\mathbf{53.73}_{\pm0.30}$ & $\mathbf{69.06}_{\pm0.25}$ & $\underline{78.34}_{\pm0.31}$ & $\color{blue!60!green}{8.19_{\pm0.07}}$ & $20.01_{\pm0.14}$ \\
            \Xhline{1pt}
        \end{tabularx}
        \label{tab:mimic3_results}
    \end{table*}

    \begin{table*}[htb]
        \centering
        \caption{Performance comparison on \textbf{MIMIC-IV}. The ground-truth DDI rate is 7.24\%. Other notations are consistent with Tab.~\ref{tab:mimic3_results}.}
        \begin{tabularx}{\textwidth}{>{\centering\arraybackslash}X|
                                    >{\centering\arraybackslash}X
                                    >{\centering\arraybackslash}X
                                    >{\centering\arraybackslash}X
                                    >{\centering\arraybackslash}X
                                    >{\centering\arraybackslash}X}
            \Xhline{1pt}
            \textbf{Model} & \textbf{Jaccard}$\uparrow$ & \textbf{F1}$\uparrow$ & \textbf{PRAUC}$\uparrow$ & \textbf{DDI}$\downarrow$ & \textbf{\# Med.}$\downarrow$ \\
            \Xhline{0.5pt}
            LR              & $47.65_{\pm0.20}$ & $63.31_{\pm0.20}$ & $74.08_{\pm0.22}$ & $\color{blue!60!green}{7.07_{\pm0.09}}$ & $13.85_{\pm0.09}$ \\
            ECC             & $45.82_{\pm0.16}$ & $61.39_{\pm0.15}$ & $73.58_{\pm0.14}$ & $\color{blue!60!green}{7.11_{\pm0.14}}$ & $13.08_{\pm0.14}$ \\
            LEAP            & $43.73_{\pm0.26}$ & $59.84_{\pm0.25}$ & $63.73_{\pm0.26}$ & $\color{blue!60!green}{6.57_{\pm0.06}}$ & $17.34_{\pm0.07}$ \\
            \Xhline{0.2pt}
            RETAIN          & $44.10_{\pm0.49}$ & $60.23_{\pm0.50}$ & $71.60_{\pm0.40}$ & $\color{orange!50!red}{7.29_{\pm0.14}}$ & $14.43_{\pm0.17}$ \\
            GAMENet         & $47.80_{\pm0.26}$ & $63.56_{\pm0.28}$ & $75.09_{\pm0.20}$ & $\color{orange!50!red}{7.85_{\pm0.05}}$ & $23.36_{\pm0.17}$ \\
            SafeDrug        & $48.47_{\pm0.17}$ & $64.24_{\pm0.17}$ & $74.27_{\pm0.22}$ & $\color{blue!60!green}{6.11_{\pm0.11}}$ & $18.15_{\pm0.11}$ \\
            DAPSNet        & $49.14_{\pm0.22}$ & $64.85_{\pm0.19}$ & $74.07_{\pm0.14}$ & $\color{blue!60!green}{5.95_{\pm0.07}}$ & $18.62_{\pm0.15}$ \\
            PROMISE        & $47.29_{\pm0.30}$ & $63.29_{\pm0.28}$ & $\mathbf{76.76}_{\pm0.28}$ & $\color{blue!60!green}{7.08_{\pm0.13}}$ & $11.34_{\pm0.08}$ \\
            COGNet          & $50.38_{\pm0.27}$ & $65.87_{\pm0.23}$ & $74.99_{\pm0.15}$ & $\color{orange!50!red}{7.79_{\pm0.15}}$ & $25.43_{\pm0.19}$ \\
            Carmen          & $50.06_{\pm0.12}$ & $65.69_{\pm0.07}$ & $74.62_{\pm0.30}$ & -- & -- \\
            DGCL            & $49.98_{\pm0.13}$ & $65.62_{\pm0.12}$ & $75.52_{\pm0.14}$ & $\color{orange!50!red}{7.50_{\pm0.04}}$ & $18.41_{\pm0.07}$ \\
            VITA            & $50.31_{\pm0.29}$ & $65.75_{\pm0.26}$ & $74.80_{\pm0.20}$ & $\color{orange!50!red}{7.45_{\pm0.11}}$ & $26.37_{\pm0.18}$ \\
            MoleRec         & $\underline{50.64}_{\pm0.23}$ & $\underline{66.23}_{\pm0.21}$ & ${75.14}_{\pm0.32}$ & $\color{blue!60!green}{6.78_{\pm0.07}}$ & $17.74_{\pm0.11}$ \\
            \Xhline{0.2pt}
            \Model          & $\mathbf{51.05}_{\pm0.28}$ & $\mathbf{66.50}_{\pm0.25}$ & $\underline{76.23}_{\pm0.28}$ & $\color{blue!60!green}{6.65_{\pm0.11}}$ & $16.53_{\pm0.16}$ \\
            \Xhline{1pt}
        \end{tabularx}
        \label{tab:mimic4_results}
    \end{table*}

    \begin{table*}[htb]
        \centering
        \caption{{Performance comparison on \textbf{eICU}. There is no DDI information available in the eICU dataset. Other notations are consistent with Tab.~\ref{tab:mimic3_results}.}}
        \begin{tabularx}{\textwidth}{>{\centering\arraybackslash}X|
                                    >{\centering\arraybackslash}X
                                    >{\centering\arraybackslash}X
                                    >{\centering\arraybackslash}X
                                    >{\centering\arraybackslash}X}
            \Xhline{1pt}
            \textbf{Model} & \textbf{Jaccard}$\uparrow$ & \textbf{F1}$\uparrow$ & \textbf{PRAUC}$\uparrow$ & \textbf{\# Med.}$\downarrow$ \\
            \Xhline{0.5pt}
            LR              & $39.96_{\pm0.22}$ & $55.33_{\pm0.21}$ & $\underline{73.24}_{\pm0.36}$ & $7.55_{\pm0.07}$ \\
            ECC             & $33.41_{\pm0.26}$ & $47.66_{\pm0.29}$ & ${72.66}_{\pm0.28}$ & $5.62_{\pm0.05}$ \\
            LEAP            & $41.25_{\pm0.18}$ & $56.43_{\pm0.21}$ & $62.67_{\pm0.29}$ & $13.36_{\pm0.04}$ \\
            RETAIN          & $41.29_{\pm0.35}$ & $56.53_{\pm0.34}$ & $71.02_{\pm0.35}$ & $9.81_{\pm0.06}$ \\
            GAMENet         & ${40.80}_{\pm0.17}$ & $56.00_{\pm0.18}$ & $68.86_{\pm0.18}$ & $12.45_{\pm0.03}$ \\
            SafeDrug        & $41.19_{\pm0.23}$ & $56.42_{\pm0.21}$ & $69.79_{\pm0.30}$ & $10.68_{\pm0.06}$ \\
            DAPSNet         & $41.14_{\pm0.28}$ & $56.44_{\pm0.30}$ & $69.99_{\pm0.31}$ & $10.87_{\pm0.06}$ \\
            PROMISE         & $40.89_{\pm0.25}$ & $56.41_{\pm0.26}$ & $\mathbf{73.54}_{\pm0.32}$ & $7.16_{\pm0.05}$ \\
            COGNet          & $41.19_{\pm0.42}$ & $56.20_{\pm0.43}$ & $67.68_{\pm0.41}$ & $20.14_{\pm0.10}$ \\
            DGCL            & $38.23_{\pm0.14}$ & $53.38_{\pm0.15}$ & $69.63_{\pm0.13}$ & $8.70_{\pm0.02}$ \\
            VITA            & $\underline{41.46}_{\pm0.26}$ & ${56.38}_{\pm0.29}$ & $67.39_{\pm0.37}$ & $18.47_{\pm0.03}$ \\
            MoleRec         & ${41.32}_{\pm0.20}$ & $\underline{56.63}_{\pm0.24}$ & ${70.14}_{\pm0.27}$ & $10.44_{\pm0.05}$ \\
            \Xhline{0.2pt}
            \Model          & $\mathbf{41.95}_{\pm0.23}$ & $\mathbf{56.97}_{\pm0.23}$ & ${68.71}_{\pm0.27}$ & $16.32_{\pm0.02}$ \\
            \Xhline{1pt}
        \end{tabularx}
        \label{tab:eicu_results}
    \end{table*}

    \subsection{Performance Comparison}
    As shown in Tab.~\ref{tab:mimic3_results}, Tab.~\ref{tab:mimic4_results}, and Tab.~\ref{tab:eicu_results}, 
    \Model~achieves the best or highly competitive overall performance across the three benchmark datasets (\textbf{MIMIC-III}, \textbf{MIMIC-IV}, and \textbf{eICU}), particularly on Jaccard and F1, while maintaining favorable safety/compactness trade-offs, demonstrating strong generalization ability across diverse clinical settings.

    \textbf{Machine learning (ML) methods} (LR and ECC) achieve relatively low performance, as they rely on simple multi-hot encodings and fail to capture temporal or relational dependencies among medical entities.  
    \textbf{Instance-based methods} (e.g., LEAP) make predictions solely based on the current visit, ignoring both patient history and inter-visit references from similar visits, which further limits their accuracy.  
    \textbf{Longitudinal methods} (e.g., RETAIN, GAMENet, SafeDrug, COGNet, DGCL, and MoleRec) perform substantially better by modeling temporal dependencies within patient histories.  
    Among them, \textbf{SafeDrug} effectively reduces DDI rates by incorporating drug–drug interaction information, though at the cost of slightly lower accuracy.  
    \textbf{COGNet} employs sequence generation with beam search, achieving modest accuracy gains but increasing the number of recommended medications and thus DDI risk.  
    \textbf{VITA} retrieves relevant past visits via target-aware attention, yet its performance varies with dataset temporal structures.  
    \textbf{Carmen}, \textbf{DGCL}, and \textbf{MoleRec} further integrate co-occurrence relationships, contrastive learning, and molecular structures, respectively, to enhance medication representations.  
    However, these models mainly focus on intra-patient historical modeling and overlook inter-visit reference augmentation via visit-conditioned retrieval.

    {Compared with \textbf{DAPSNet} and \textbf{PROMISE}, the performance improvements of \Model\ primarily stem from its \emph{representation-consistent} design that couples representation learning and visit-conditioned inter-visit retrieval.
    Although DAPSNet and PROMISE also incorporate similarity information to improve recommendation accuracy, their retrieval mechanisms are \emph{decoupled} from representation learning: DAPSNet relies on embedding-level similarity and answer-side matching, while PROMISE computes trajectory similarity via DTW followed by hierarchical attention. 
    This separation can lead to a mismatch between the learned embedding space and the retrieval criterion. 
    In contrast, \Model\ performs retrieval and aggregation within the same hypergraph representation space, enabling coherent use of inter-visit references to refine latent condition estimation and medication prediction.
    This representation–retrieval consistency ensures semantically aligned visit matching and contributes to the observed performance gains across multiple evaluation metrics.}

    Consequently, \Model~surpasses the strongest baselines on the primary recommendation metrics across all three datasets, achieving Jaccard score improvements of $+0.61\%$, $+0.41\%$, and $+0.49\%$ on MIMIC-III, MIMIC-IV, and eICU, respectively.  
    Moreover, \Model~maintains DDI rates below or comparable to real-world prescriptions (8.68\% for MIMIC-III and 7.24\% for MIMIC-IV), indicating that its superior accuracy does not compromise safety (on MIMIC-III/IV where DDI graphs are available).  
    {Notably, the results on the \textbf{eICU} dataset further demonstrate \Model’s robustness and generalization capability on cross-domain, cross-institution, and cross-population datasets.}  
    Overall, \Model~achieves a favorable balance between accuracy and safety, underscoring its potential as a reliable clinical decision-support system.

    \begin{table*}[htb]
    \centering
    \caption{
    Ablation study on \textbf{MIMIC-III} and \textbf{MIMIC-IV}. 
    Results are reported for both \MedRep\ and \MedRec\ ablations. 
    Jaccard, F1, PRAUC, and DDI are expressed as percentages (\%), while \#Med.\ retains its original scale. 
    Values denote mean $\pm$ standard deviation over ten bootstrap sampling. 
    }
    \begin{tabularx}{\textwidth}{
    >{\raggedright\arraybackslash}p{3cm} |
    *{5}{>{\centering\arraybackslash}X}}
    \Xhline{1pt}
    \multicolumn{6}{c}{\textbf{MIMIC-III}} \\
    \Xhline{0.5pt}
    \textbf{Model} & \textbf{Jaccard}$\uparrow$ & \textbf{F1}$\uparrow$ & \textbf{PRAUC}$\uparrow$ & \textbf{DDI}$\downarrow$ & \textbf{\#Med.}$\downarrow$ \\
    \Xhline{0.5pt}
    \Model\                & $\mathbf{53.10_{\pm0.43}}$ & $\mathbf{68.49_{\pm0.37}}$ & $77.32_{\pm0.37}$ & $6.67_{\pm0.06}$ & $22.77_{\pm0.13}$ \\
    \quad– \MedRec– w/o Sim.        & $52.91_{\pm0.42}$ & $68.33_{\pm0.37}$ & $\mathbf{77.44_{\pm0.34}}$ & $6.40_{\pm0.06}$ & $22.50_{\pm0.13}$ \\
    \quad– \MedRec– w/o Hist.       & $51.90_{\pm0.41}$ & $67.40_{\pm0.37}$ & $76.40_{\pm0.44}$ & $6.12_{\pm0.06}$ & $23.16_{\pm0.17}$ \\
    \quad– \MedRep–None     & $51.70_{\pm0.46}$ & $67.23_{\pm0.42}$ & $76.28_{\pm0.36}$ & $6.52_{\pm0.06}$ & $23.51_{\pm0.14}$ \\
    \quad{– \MedRep–Fixed}    & $50.49_{\pm0.23}$ & $66.24_{\pm0.21}$ & $75.44_{\pm0.33}$ & $6.45_{\pm0.04}$ & $22.74_{\pm0.13}$ \\
    \quad– \MedRep–HGCN     & $52.17_{\pm0.45}$ & $67.67_{\pm0.40}$ & $75.66_{\pm0.35}$ & $6.01_{\pm0.05}$ & $22.37_{\pm0.13}$ \\
    \quad– \MedRep–GCN      & $51.40_{\pm0.40}$ & $66.96_{\pm0.36}$ & $76.31_{\pm0.38}$ & $6.05_{\pm0.05}$ & $23.50_{\pm0.12}$ \\
    \Xhline{1pt}
    \end{tabularx}

    \vspace{0.5cm}

    \begin{tabularx}{\textwidth}{
    >{\raggedright\arraybackslash}p{3cm} |
    *{5}{>{\centering\arraybackslash}X}}
    \Xhline{1pt}
    \multicolumn{6}{c}{\textbf{MIMIC-IV}} \\
    \Xhline{0.5pt}
    \textbf{Model} & \textbf{Jaccard}$\uparrow$ & \textbf{F1}$\uparrow$ & \textbf{PRAUC}$\uparrow$ & \textbf{DDI}$\downarrow$ & \textbf{\#Med.}$\downarrow$ \\
    \Xhline{0.5pt}
    \Model\                & $\mathbf{50.92_{\pm0.25}}$ & $\mathbf{66.35_{\pm0.24}}$ & $\mathbf{75.96_{\pm0.26}}$ & $6.60_{\pm0.13}$ & $16.68_{\pm0.11}$ \\
    \quad– \MedRec–w/o Sim.        & $50.52_{\pm0.26}$ & $66.03_{\pm0.23}$ & $75.91_{\pm0.24}$ & $6.79_{\pm0.15}$ & $16.61_{\pm0.12}$ \\
    \quad– \MedRec–w/o Hist.       & $49.98_{\pm0.23}$ & $65.52_{\pm0.21}$ & $75.19_{\pm0.25}$ & $6.47_{\pm0.11}$ & $16.26_{\pm0.11}$ \\
    \quad– \MedRep–None     & $50.15_{\pm0.25}$ & $65.68_{\pm0.23}$ & $75.47_{\pm0.25}$ & $6.65_{\pm0.11}$ & $16.61_{\pm0.11}$ \\
    \quad{– \MedRep–Fixed}    & $46.68_{\pm0.21}$ & $62.58_{\pm0.22}$ & $72.99_{\pm0.27}$ & $6.01_{\pm0.07}$ & $21.11_{\pm0.12}$ \\
    \quad– \MedRep–HGCN     & $50.21_{\pm0.23}$ & $65.76_{\pm0.22}$ & $75.54_{\pm0.30}$ & $6.48_{\pm0.12}$ & $16.63_{\pm0.12}$ \\
    \quad– \MedRep–GCN      & $50.25_{\pm0.20}$ & $65.77_{\pm0.19}$ & $75.61_{\pm0.23}$ & $6.48_{\pm0.12}$ & $16.95_{\pm0.11}$ \\
    \Xhline{1pt}
    \end{tabularx}
    \label{tab:ablation_main}
    \end{table*}
    
    \subsection{Ablation Study}
    To evaluate the contribution of each component in \Model, we conduct comprehensive ablation studies on the \textbf{MIMIC-III}~\cite{MIMIC-III} and \textbf{MIMIC-IV}~\cite{MIMIC-IV} datasets.
    We construct several model variants as follows:
    (1) \Model–\MedRep–None initializes entity and visit representations randomly, removing the pretrained entity and visit representations from \MedRep;
    (2) {\Model–\MedRep–Fixed keeps the embeddings generated by \MedRep~frozen during \MedRec~training, preventing task-specific adaptation;}
    (3) \Model–\MedRep–HGCN replaces the KHGE encoder in \MedRep~with a standard HGCN~\cite{hgnn};
    (4) \Model–\MedRep–GCN replaces the KHGE encoder with a conventional GCN~\cite{GNN};
    (5) \Model–\MedRec–w/o Hist. removes the historical information encoder, relying solely on the similarity channel; and
    (6) \Model–\MedRec–w/o Sim. disables the similarity channel, preserving only the historical encoder.

    From the results, several key observations can be drawn.  
    First, all ablated variants show degraded performance compared to the full \Model, confirming the overall effectiveness and complementarity of its components.  
    Second, removing \MedRep~(\Model–\MedRep–None) leads to a notable decline (e.g., Jaccard decreases from $53.10 \to 51.70$ in MIMIC-III), highlighting the essential role of pretrained representations in encoding medical knowledge.  
    {Notably, the \Model–\MedRep–Fixed variant, which freezes the pretrained embeddings without further optimization, exhibits a more pronounced decrease compared to the jointly optimized model.
    This confirms that fine-tuning the pretrained representations during \MedRec~training is crucial for aligning them with the downstream recommendation objective.}
    Furthermore, substituting the KHGE encoder with HGCN or GCN (\Model–\MedRep–HGCN / \Model–\MedRep–GCN) produces only marginal gains (Jaccard: $50.15\to50.25$ in MIMIC-IV) over the random baseline, indicating that conventional graph encoders are less capable of capturing intra-visit combinatorial semantics among medications.  
    Finally, removing the historical channel (\Model–\MedRec–w/o Hist.) results in the largest degradation (e.g., F1: $68.49\to67.40$ / $66.35\to65.52$ in MIMIC-III/IV respectively), underscoring the critical importance of longitudinal information. 
    Nevertheless, the two channels are complementary, and their integration yields the best overall performance.


    \subsection{Hyperparameters Analysis}
    In \Model, three hyperparameters potentially influence the final performance: 
    (1) the number of pretraining epochs in the \MedRep~stage, 
    (2) the number of retrieved similar visits (Top-$n$) in \MedRec, 
    and (3) the number of augmented hypergraph views used for contrastive learning. 
    We conduct a comprehensive sensitivity analysis to evaluate their effects, 
    with results summarized in Tab.~\ref{tab:pretrain_sensitivity}, 
    Tab.~\ref{tab:topn_sensitivity}, and Tab.~\ref{tab:views_sensitivity}. 
    All experiments evaluate five metrics—Jaccard, F1, PRAUC, DDI rate, and average number of medications—while keeping all other settings fixed.

    \begin{table*}[htb]
\centering
\caption{
Pretraining-epoch sensitivity analysis on \textbf{MIMIC-III} and \textbf{MIMIC-IV}. 
Jaccard, F1, PRAUC, and DDI are reported in percentage (\%), while \#Med.\ retains its original scale.
Values denote mean $\pm$ standard deviation over ten bootstrap sampling. 
}
\begin{tabularx}{\textwidth}{
  >{\raggedright\arraybackslash}p{2cm} |
  >{\centering\arraybackslash}p{1.8cm} |
  *{5}{>{\centering\arraybackslash}X}}
\toprule
\textbf{Dataset} & \textbf{Epoch} & \textbf{Jaccard}$\uparrow$ & \textbf{F1}$\uparrow$ & \textbf{PRAUC}$\uparrow$ & \textbf{DDI}$\downarrow$ & \textbf{\#Med.}$\downarrow$ \\
\midrule
\multirow{5}{*}{\textbf{MIMIC-III}} 
 & 0    & $52.27_{\pm0.25}$ & $67.75_{\pm0.22}$ & $\mathbf{77.08}_{\pm0.32}$ & $7.93_{\pm0.10}$ & $21.90_{\pm0.17}$ \\
 & 100  & $\mathbf{52.48}_{\pm0.23}$ & $\mathbf{67.97}_{\pm0.20}$ & $76.10_{\pm0.29}$ & $6.00_{\pm0.04}$ & $21.93_{\pm0.10}$ \\
 & 300  & $52.26_{\pm0.25}$ & $67.78_{\pm0.22}$ & $76.27_{\pm0.39}$ & $6.06_{\pm0.04}$ & $22.74_{\pm0.11}$ \\
 & 1000 & $51.91_{\pm0.22}$ & $67.41_{\pm0.19}$ & $75.95_{\pm0.37}$ & $6.28_{\pm0.04}$ & $22.95_{\pm0.14}$ \\
\midrule
\multirow{5}{*}{\textbf{MIMIC-IV}} 
 & 0    & $49.87_{\pm0.23}$ & $65.48_{\pm0.23}$ & $75.60_{\pm0.27}$ & $6.30_{\pm0.10}$ & $16.50_{\pm0.14}$ \\
 & 100  & $50.03_{\pm0.23}$ & $65.65_{\pm0.22}$ & $75.87_{\pm0.27}$ & $6.34_{\pm0.07}$ & $16.81_{\pm0.17}$ \\
 & 300  & $\mathbf{50.17}_{\pm0.33}$ & $\mathbf{65.75}_{\pm0.32}$ & $\mathbf{75.92}_{\pm0.27}$ & $6.39_{\pm0.09}$ & $16.32_{\pm0.15}$ \\
 & 1000 & $49.66_{\pm0.26}$ & $65.30_{\pm0.26}$ & $75.67_{\pm0.25}$ & $6.46_{\pm0.09}$ & $16.20_{\pm0.16}$ \\
\bottomrule
\end{tabularx}
\label{tab:pretrain_sensitivity}
\end{table*}

    \noindent\textbf{(1) Number of Pretraining Epochs.} 
    We vary the number of pretraining epochs over $\{0, 100, 300, 1000\}$ 
    to examine the effect of contrastive pretraining duration. 
    As shown in Tab.~\ref{tab:pretrain_sensitivity}, performance on both datasets first improves and then slightly declines as the number of epochs increases, 
    indicating that moderate pretraining enhances representation quality, 
    whereas excessive training may lead to overfitting or representation collapse.

\begin{table*}[htb]
\centering
\caption{
Top-$n$ sensitivity analysis on \textbf{MIMIC-III} and \textbf{MIMIC-IV}. 
Jaccard, F1, PRAUC, and DDI are expressed as percentages (\%), while \#Med.\ retains its original scale. 
Values denote mean $\pm$ standard deviation over ten bootstrap sampling. 
}
\begin{tabularx}{\textwidth}{
  >{\raggedright\arraybackslash}p{2cm} |
  >{\centering\arraybackslash}p{1.2cm} |
  *{5}{>{\centering\arraybackslash}X}}
\toprule
\textbf{Dataset} & \textbf{Top-$n$} & \textbf{Jaccard}$\uparrow$ & \textbf{F1}$\uparrow$ & \textbf{PRAUC}$\uparrow$ & \textbf{DDI}$\downarrow$ & \textbf{\#Med.}$\downarrow$ \\
\midrule
\multirow{6}{*}{\textbf{MIMIC-III}} 
 & 0    & $53.10_{\pm0.43}$ & $68.49_{\pm0.37}$ & $77.32_{\pm0.37}$ & $6.67_{\pm0.06}$ & $22.77_{\pm0.13}$ \\
 & 1    & $\mathbf{53.23}_{\pm0.40}$ & $\mathbf{68.59}_{\pm0.35}$ & $\mathbf{77.50}_{\pm0.35}$ & $6.52_{\pm0.04}$ & $22.58_{\pm0.13}$ \\
 & 10   & $52.93_{\pm0.37}$ & $68.33_{\pm0.33}$ & $77.43_{\pm0.38}$ & $6.25_{\pm0.05}$ & $22.63_{\pm0.13}$ \\
 & 100  & $52.88_{\pm0.40}$ & $68.27_{\pm0.35}$ & $77.20_{\pm0.36}$ & $6.44_{\pm0.06}$ & $23.31_{\pm0.13}$ \\
 & 500  & $52.80_{\pm0.38}$ & $68.22_{\pm0.34}$ & $77.13_{\pm0.36}$ & $6.39_{\pm0.06}$ & $23.20_{\pm0.12}$ \\
 & 1000 & $52.86_{\pm0.37}$ & $68.26_{\pm0.33}$ & $77.16_{\pm0.39}$ & $6.40_{\pm0.05}$ & $22.97_{\pm0.12}$ \\
\midrule
\multirow{6}{*}{\textbf{MIMIC-IV}} 
 & 0    & $50.52_{\pm0.26}$ & $66.03_{\pm0.23}$ & $75.91_{\pm0.24}$ & $6.79_{\pm0.15}$ & $16.61_{\pm0.12}$ \\
 & 1    & $50.51_{\pm0.25}$ & $66.01_{\pm0.24}$ & $75.79_{\pm0.21}$ & $6.69_{\pm0.12}$ & $16.93_{\pm0.11}$ \\
 & 10   & $50.39_{\pm0.23}$ & $65.90_{\pm0.21}$ & $75.98_{\pm0.23}$ & $6.49_{\pm0.12}$ & $16.69_{\pm0.11}$ \\
 & 100  & $50.70_{\pm0.18}$ & $\mathbf{66.23}_{\pm0.17}$ & $76.18_{\pm0.23}$ & $6.49_{\pm0.13}$ & $16.80_{\pm0.12}$ \\
 & 500  & $\mathbf{50.77}_{\pm0.21}$ & $66.29_{\pm0.19}$ & $\mathbf{76.30}_{\pm0.23}$ & $6.52_{\pm0.13}$ & $16.51_{\pm0.12}$ \\
 & 1000 & $50.72_{\pm0.15}$ & $66.22_{\pm0.14}$ & $76.25_{\pm0.22}$ & $6.54_{\pm0.10}$ & $16.42_{\pm0.12}$ \\
\bottomrule
\end{tabularx}
\label{tab:topn_sensitivity}
\end{table*}

    \noindent\textbf{(2) Number of Retrieved Similar Visits.} 
    To assess the impact of inter-visit reference retrieval, we vary the number of retrieved similar visits (Top-$n$) in $\{0, 1, 10, 100, 500, 1000\}$. 
    As shown in Tab.~\ref{tab:topn_sensitivity}, incorporating a small number of similar visits significantly enhances predictive performance, 
    whereas a large $n$ introduces noise or less relevant samples, leading to marginal declines in Jaccard and F1. 
    Nevertheless, the DDI rate and average medication count remain stable, suggesting that \Model~achieves a robust balance between accuracy and safety.

\begin{table*}[htb]
\centering
\caption{{Performance and GPU memory usage under different numbers of views on \textbf{MIMIC-III} and \textbf{MIMIC-IV}. 
Jaccard, F1, PRAUC, and DDI are reported in percentage (\%); \#Med.\ and GPU Memory (GiB) keep their original scales.}}
\begin{tabularx}{\textwidth}{
  >{\raggedright\arraybackslash}p{1.5cm} |
  >{\centering\arraybackslash}p{1cm} |
  *{6}{>{\centering\arraybackslash}X}}
\toprule
\textbf{Dataset} & \textbf{\#Views} & \textbf{Jaccard}$\uparrow$ & \textbf{F1}$\uparrow$ & \textbf{PRAUC}$\uparrow$ & \textbf{DDI}$\downarrow$ & \textbf{\#Med.}$\downarrow$ & \textbf{GPU Mem.} \\
\midrule
\multirow{3}{*}{\textbf{MIMIC-III}} 
 & $2$ & $51.13_{\pm0.17}$ & $66.77_{\pm0.16}$ & $75.73_{\pm0.34}$ & $6.50_{\pm0.05}$ & $23.80_{\pm0.14}$ & $9.76$ \\
 & $3$ & $51.26_{\pm0.22}$ & $66.85_{\pm0.18}$ & $75.86_{\pm0.40}$ & $6.23_{\pm0.05}$ & $24.60_{\pm0.12}$ & $14.09$ \\
 & $4$ & $51.05_{\pm0.20}$ & $66.67_{\pm0.19}$ & $75.89_{\pm0.43}$ & $6.33_{\pm0.04}$ & $23.49_{\pm0.12}$ & $24.53$ \\
\midrule
\multirow{3}{*}{\textbf{MIMIC-IV}} 
 & $2$ & $48.67_{\pm0.23}$ & $64.40_{\pm0.22}$ & $73.57_{\pm0.32}$ & $5.90_{\pm0.07}$ & $20.93_{\pm0.17}$ & $14.46$ \\
 & $3$ & $48.61_{\pm0.21}$ & $64.34_{\pm0.20}$ & $73.44_{\pm0.27}$ & $5.99_{\pm0.06}$ & $21.19_{\pm0.18}$ & $23.75$ \\
 & $4$ & $48.66_{\pm0.25}$ & $64.39_{\pm0.25}$ & $73.31_{\pm0.29}$ & $6.03_{\pm0.07}$ & $20.87_{\pm0.17}$ & $38.61$ \\
\bottomrule
\end{tabularx}
\label{tab:views_sensitivity}
\end{table*}

    \noindent {\textbf{(3) Number of Augmented Views.} 
    In lines~5--6 of Alg.~\ref{alg:training_pipeline}, \Model~generates two augmented hypergraph views for contrastive pretraining in the \MedRep~stage. 
    This design follows standard contrastive frameworks~\cite{TriCL, SGL}, where two correlated views are typically sufficient to learn discriminative yet invariant representations. 
    To validate this choice, we vary the number of augmented views over $\{2, 3, 4\}$. 
    As shown in Tab.~\ref{tab:views_sensitivity}, adding more views provides negligible improvements or even slight degradation, 
    while GPU memory usage increases sharply (from 9.8$\to$24.5~GiB on MIMIC-III and 14.5$\to$38.6~GiB on MIMIC-IV). 
    These findings suggest that two views strike the best balance between performance and computational efficiency, 
    supporting the design adopted in Alg.~\ref{alg:training_pipeline}.}

    \begin{figure}[htb]
        \centering
        \subfloat[MIMIC-III]{
            \includegraphics[width=0.45\textwidth]{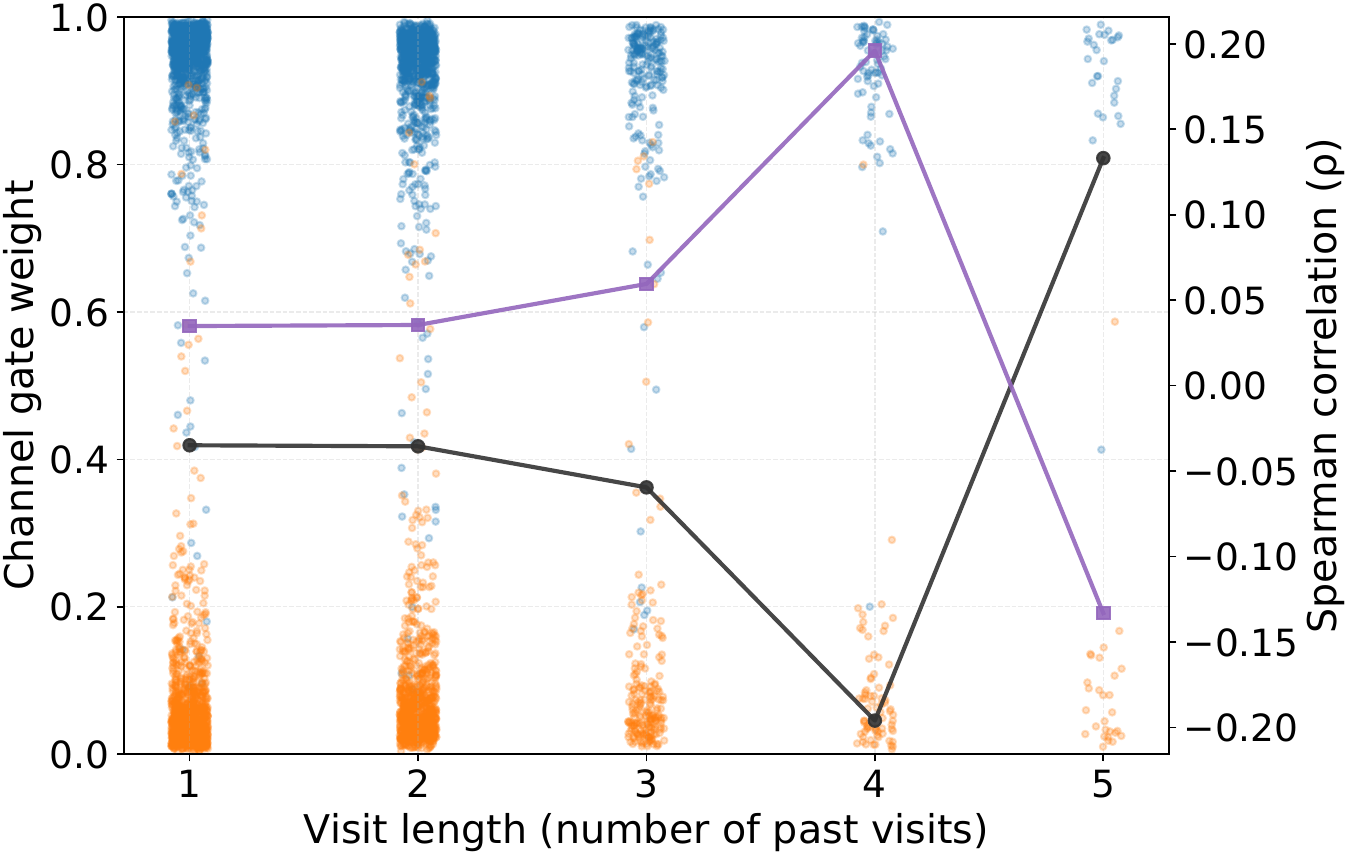}
            \label{fig:sub1}
        }
        \subfloat[MIMIC-IV]{
            \includegraphics[width=0.45\textwidth]{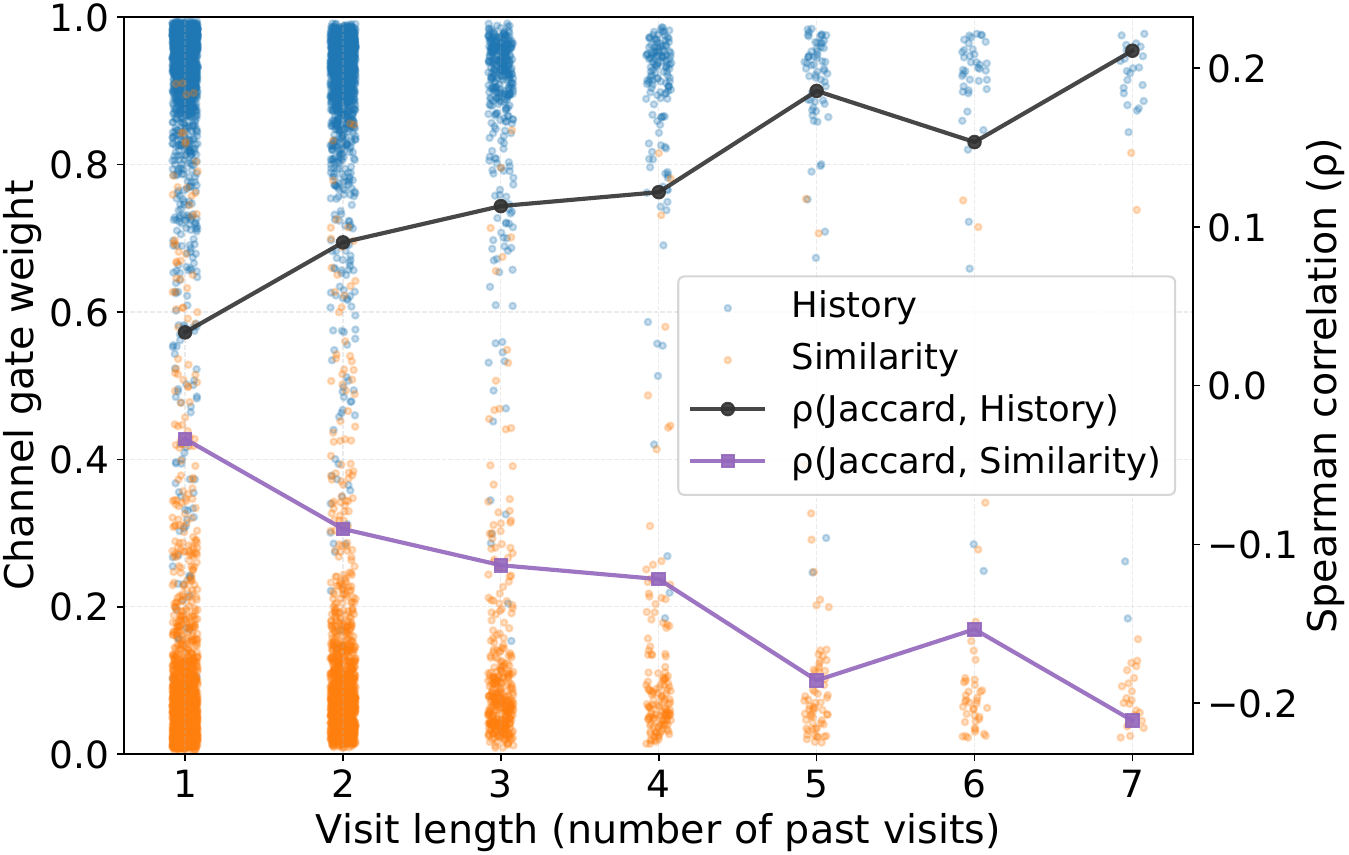}
            \label{fig:sub2}
        }

        \caption{
        {Visualization of adaptive channel weighting and its association with model performance across different visit lengths. 
        Blue and orange dots represent per-visit gate weights for the history and similarity channels, respectively. 
        Black and purple lines indicate the Spearman correlations ($\rho$) between Jaccard performance and each channel’s weight, computed within visit-length groups. 
        To ensure statistical reliability, visit-length groups with fewer than 20 samples were excluded.}
        }
        \label{fig:combined}
    \end{figure}

    \subsection{Interpretability analysis}
 
    {To better understand how \Model\ balances information from different sources, we visualize the per-visit gate weights of the two channels learned by the fusion module (blue: history channel; orange: similarity channel) in Fig.~\ref{fig:combined}, along with their Spearman correlations ($\rho$) with predictive performance (Jaccard). Each dot represents a single visit.}

    {In both datasets, the similarity channel generally receives smaller gate weights than the history channel, 
    since the latter always incorporates the current visit’s diagnosis and procedure information—even for single-visit patients—whereas the similarity channel depends on retrieved visits and provides supplementary context when longitudinal data are sparse. 
    This finding aligns with our ablation results, where removing the history channel led to a larger performance degradation.  
    However, the strength of association between each channel and predictive accuracy varies across datasets.  
    In MIMIC-III, where patient trajectories are shorter and sparser, the similarity-channel weight correlates positively with performance during early visits (1–4), suggesting that retrieved visits compensate for limited historical context.  
    Once sufficient history accumulates ($\geq$5 visits), the correlation shifts toward the history channel, indicating increasing reliance on personal longitudinal information.  
    In contrast, MIMIC-IV contains richer and longer patient trajectories, and its performance correlates more strongly with the history channel across all visit lengths.}

    {Overall, the correlation between predictive accuracy and the similarity-channel weight exhibits a general downward trend as visit length increases, 
    indicating that \Model\ gradually shifts its reliance from retrieved similar visits to accumulated patient-specific history.  
    This visualization thus provides an intrinsic, model-level diagnostic perspective on how the relative contributions of different information sources evolve with increasing data completeness.}

    \begin{table*}[htb]
    \centering
    \caption{
    {Performance comparison under the cold-start (first-visit) setting on \textbf{MIMIC-III} and \textbf{MIMIC-IV}. 
    All metrics are reported as percentages (\%), except for the average number of medications (\#Med.). 
    Values denote mean $\pm$ standard deviation over ten bootstrapped runs. 
    A total of 192 and 307 patients have only one visit in MIMIC-III and MIMIC-IV, respectively.}
    }
    \begin{tabularx}{\textwidth}{
    >{\raggedright\arraybackslash}p{1.6cm} |
    >{\centering\arraybackslash}p{4cm} |
    *{5}{>{\centering\arraybackslash}X}}
    \toprule
    \textbf{Dataset} & \textbf{Model} & \textbf{Jaccard}$\uparrow$ & \textbf{F1}$\uparrow$ & \textbf{PRAUC}$\uparrow$ & \textbf{DDI}$\downarrow$ & \textbf{\#Med.}$\downarrow$ \\
    \midrule
    \multirow{6}{*}{\textbf{MIMIC-III}} 
    & SafeDrug            & $49.47_{\pm1.24}$ & $65.22_{\pm1.03}$ & $74.92_{\pm0.91}$ & $6.22_{\pm0.20}$ & $18.01_{\pm0.26}$ \\
    & DAPSNet             & $50.68_{\pm0.95}$ & $66.20_{\pm0.79}$ & $74.90_{\pm1.00}$ & $6.12_{\pm0.11}$ & $19.92_{\pm0.33}$ \\
    & COGNet              & $51.19_{\pm1.20}$ & $66.64_{\pm1.11}$ & $75.62_{\pm0.69}$ & $8.62_{\pm0.15}$ & $24.90_{\pm0.27}$ \\
    \cmidrule(lr){2-7}  
    & \Model\ & $\mathbf{51.60}_{\pm0.86}$ & $\mathbf{67.14}_{\pm0.73}$ & $\mathbf{75.83}_{\pm0.98}$ & $8.65_{\pm0.29}$ & $20.04_{\pm0.31}$ \\
    & \Model\ -\MedRec- w/o Hist.   & $49.93_{\pm1.00}$ & $65.46_{\pm0.92}$ & $74.78_{\pm1.04}$ & $6.19_{\pm0.25}$ & $23.44_{\pm0.32}$ \\
    & \Model\ -\MedRec- w/o Sim.     & $49.52_{\pm1.08}$ & $65.24_{\pm0.98}$ & $74.62_{\pm0.96}$ & $6.35_{\pm0.25}$ & $22.43_{\pm0.28}$ \\
    \midrule
    \multirow{6}{*}{\textbf{MIMIC-IV}} 
    & SafeDrug            & $44.89_{\pm1.16}$ & $60.61_{\pm1.14}$ & $66.83_{\pm1.34}$ & $6.41_{\pm0.24}$ & $17.58_{\pm0.31}$ \\
    & DAPSNet             & $49.49_{\pm1.10}$ & $65.23_{\pm1.01}$ & $73.71_{\pm1.05}$ & $6.07_{\pm0.22}$ & $17.49_{\pm0.25}$ \\
    & COGNet              & $46.91_{\pm0.48}$ & $62.64_{\pm0.46}$ & $72.82_{\pm0.70}$ & $8.90_{\pm0.17}$ & $18.67_{\pm0.39}$ \\
    \cmidrule(lr){2-7}  
    & \Model\ & $\mathbf{50.30}_{\pm0.94}$ & $\mathbf{65.78}_{\pm0.86}$ & $\mathbf{75.19}_{\pm1.11}$ & $6.81_{\pm0.37}$ & $16.31_{\pm0.27}$ \\
    & \Model\ -\MedRec- w/o Hist.   & $49.84_{\pm0.98}$ & $65.39_{\pm0.89}$ & $74.94_{\pm1.13}$ & $6.61_{\pm0.34}$ & $16.30_{\pm0.25}$ \\
    & \Model\ -\MedRec- w/o Sim.     & $49.62_{\pm1.00}$ & $65.15_{\pm0.94}$ & $74.89_{\pm1.12}$ & $7.18_{\pm0.26}$ & $16.86_{\pm0.36}$ \\
    \bottomrule
    \end{tabularx}
    \label{tab:coldstart_results}
    \end{table*}

    \subsection{Cold-start performance analysis}
    {To evaluate whether the proposed similar-visit retrieval mechanism effectively alleviates the cold-start problem, 
    we perform a targeted assessment under the no-history (first-visit) setting, as shown in Tab.~\ref{tab:coldstart_results}. 
    Across both \textbf{MIMIC-III} and \textbf{MIMIC-IV}, \Model\ achieves the best performance among all baselines 
    (\textit{SafeDrug}, \textit{DAPSNet}, and \textit{COGNet}) in terms of Jaccard, F1, and PRAUC, 
    demonstrating strong generalization capability even when patient history is unavailable.}

    {Compared with its ablated variants, the complete model exhibits clear advantages. 
    Removing the similarity channel (\textit{w/o Sim.}) leads to consistent declines in all major metrics—on \textbf{MIMIC-III}, Jaccard decreases by approximately $2\%$ and F1 by $1.5\%$—indicating that retrieved visits provide valuable external priors that compensate for the absence of personal history. 
    In contrast, removing the historical channel (\textit{w/o Hist.}) results in slightly lower performance, suggesting that under the single-visit condition, the model mainly relies on current-visit information while still benefiting from the structural representation of the historical channel. 
    Together with the observations from Fig.~\ref{fig:combined}, these results highlight a complementary effect between the two channels: 
    the similarity channel is crucial in true cold-start cases, whereas the historical channel becomes increasingly influential as longitudinal information accumulates.}

    {These findings quantitatively verify that the dual-channel design enables \Model\ to adaptively balance retrieved population-level knowledge and patient-specific information, thereby achieving superior robustness and predictive accuracy in cold-start scenarios.}

    \subsection{Retrieval-quality correlation analysis}
\begin{figure}[htbp]
        \centering
        \includegraphics[width=\linewidth]{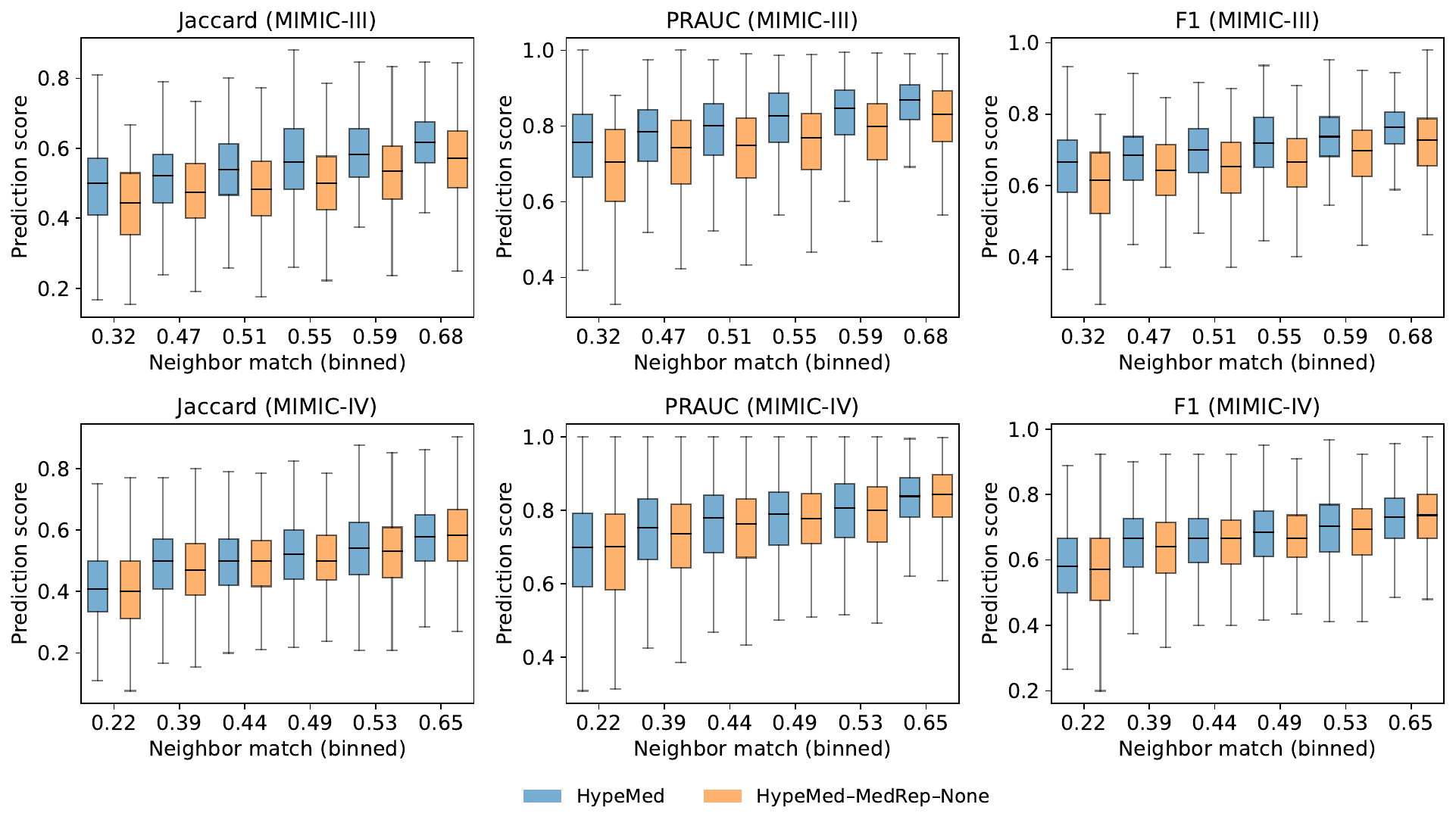}
        \caption{\rev{Retrieval-quality correlation analysis. We compare \Model\ with \Model-\MedRep-None, a controlled variant that keeps the same prediction architecture but removes \MedRep\ pre-training, resulting in retrieval within a non-pretrained embedding space. Test visits are grouped into bins by retrieval quality, measured as the F1 score between medications in the retrieved visits and the ground-truth medications. The x-axis denotes retrieval-quality bins (higher is better), and the y-axis reports the final prediction performance.}}
        \label{fig:retrieval-quality}
    \end{figure}

    \rev{To quantify the advantage of \emph{joint} representation--retrieval learning over a decoupled retrieval paradigm, we report the retrieval-quality correlation analysis in Fig.~\ref{fig:retrieval-quality}.}
    
    \rev{From Fig.~\ref{fig:retrieval-quality}, we observe two key trends:
    First, for both \Model\ and \Model-\MedRep-None, higher retrieval quality generally leads to better prediction performance, confirming that relevant references provide useful complementary signals~\cite{COGNet, VITA, DAPSNet}.
    Second, under the \emph{same} retrieval-quality bin, \Model\ consistently achieves higher accuracy than \Model-\MedRep-None, indicating that the performance gain is not solely explained by retrieving ``more correct'' cases.}
    
    \rev{This result supports our hypothesis that \emph{coupling representation learning with retrieval improves how retrieved context is \emph{represented} and \emph{used}}.
    In \Model-None, retrieval is performed in a randomly initialized or weakly regularized embedding space; even when the retrieved visits exhibit comparable label overlap, their embeddings may be poorly organized with respect to visit semantics, making the fused context less predictable and less exploitable by the downstream predictor.
    In contrast, \MedRep\ pre-trains a hyperedge-aware metric space where visit embeddings are explicitly optimized to encode high-order clinical interactions and preserve set-level semantics.
    Therefore, retrieved neighbors become not only label-relevant but also \emph{geometrically compatible} with the query visit, providing a more structured and informative signal for fusion, which ultimately translates into consistently better prediction performance.}

    \begin{figure}[htb]
        \centering
        \includegraphics[width=0.75\textwidth, keepaspectratio]{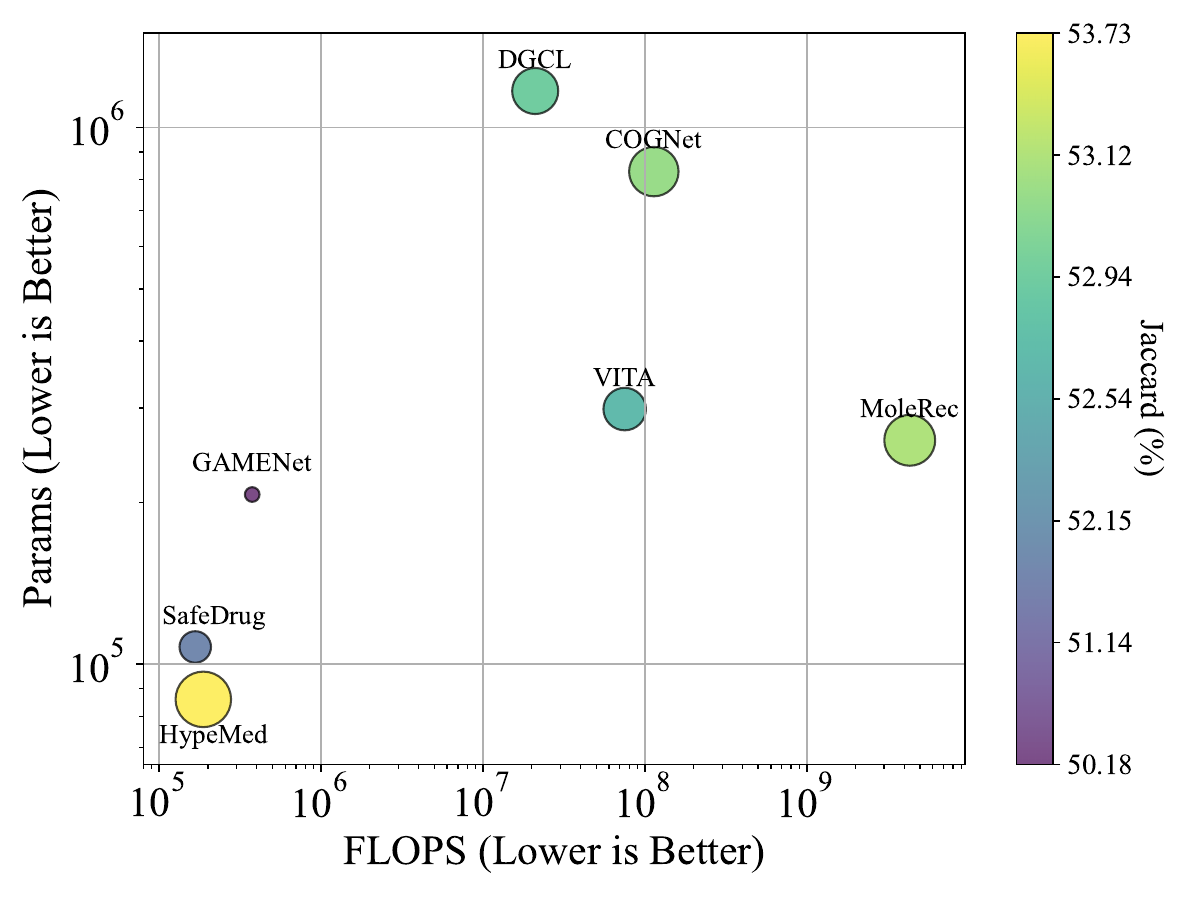}
        \caption{
            Comparative time complexity across models (FLOPs vs. Parameters)
        }
        \label{fig:flops}
    \end{figure}

    \subsection{Time complexity comparison}
    We perform a time complexity analysis by comparing our approach with state-of-the-art methods, using floating point operations (FLOPs) as the metric. Specifically, we measure the FLOPs for each model during the inference phase. The results are illustrated in Fig.~\ref{fig:flops}.
    Thanks to its two-stage design, \Model~reuses \MedRep\ embeddings and avoids additional GNN computations during inference. This significantly reduces its computational load, making its FLOPs count second only to SafeDrug. On the other hand, MoleRec executes a 4-layer GNN across multiple graphs, which results in the highest FLOPs during inference among the evaluated models.

    \subsection{Case study}
    To illustrate the effectiveness of \Model's recommendations more vividly, we conduct a case study with \textit{Patient-908} as an example. \textit{Patient-908} from the MIMIC-III test set has two medical visits. During the first visit, the patient was diagnosed with diseases such as cirrhosis, liver cancer, and chronic hepatitis. In the second visit, new conditions were added, including complications from liver transplantation, renal failure, and confirmed sepsis. Tab.~\ref{tab:case} displays detailed information. Due to space constraints, we use the ICD codes to represent diagnoses and ATC codes to represent medications.

\begin{figure}[htb]
        \centering
        \includegraphics[width=0.75\textwidth, keepaspectratio]{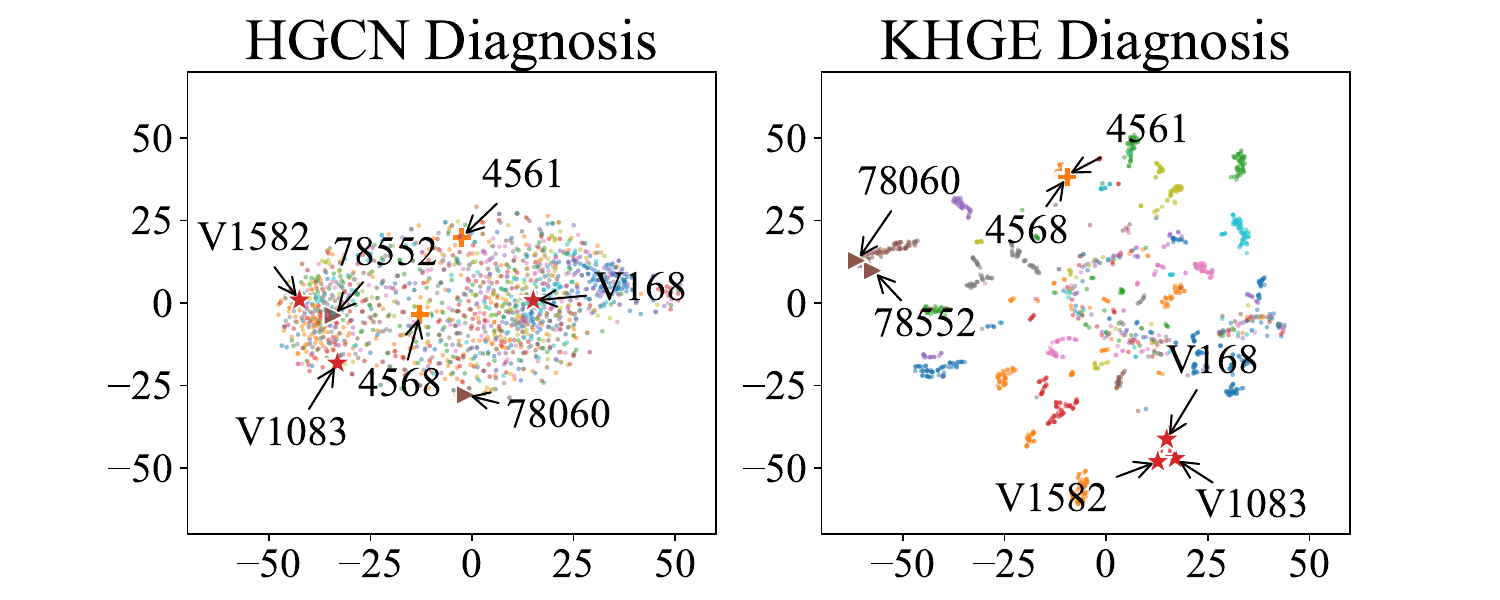}
        \caption{
            Comparative Visualization of Diagnosis Entity Representations in \MedRep~using HGCN and KHGE Encoders.
        }
        \label{fig:embed_case}
    \end{figure}

    Fig.~\ref{fig:embed_case} demonstrates the visualization of \textbf{diagnosis} entities using HGCN and KHGE as encoders in \MedRep. Points in the same color correspond to the same ICD category.
    The KHGE group embeddings show apparent clustering by ICD categories. Nodes within the same category display closer distances. In contrast, the HGCN group shows scattered distribution without clear category distinction.
    To better demonstrate the effectiveness of KHGE, we select three groups of diagnosis entities from Visit-2, each belonging to the same category in the ICD classification system. The groups are as follows: 
    \begin{itemize}
        \item ${\color{red}\star}$ \textit{Persons With Potential Health Hazards Related To Personal And Family History} (V1083, V168, and V1582);
        \item ${\color{brown}\blacktriangleright}$ \textit{Symptoms} (78060, 78552);
        \item ${\color{orange}\mathbf{+}}$ \textit{Varicose veins of other sites} (4568, 4561).
    \end{itemize}
    Different markers are used to highlight entities belonging to different groups.
    We find that in the HGCN group, the distribution of these three entity groups was dispersed, while in the KHGE group, entities belonging to the same category are clustered together.
    This suggests that by incorporating the ICD classification system into the modeling process, KHGE can translate the relative distances of entities in ICD into distances in embedding representations, granting similar representations to entities of the same category. In contrast, HGCN, lacking ICD information, fails to capture such information.

    We then compare the effectiveness of medication recommendations for \textit{Patient-908} under two scenarios: with and without similar visit information (refer to Tab.~\ref{tab:case}). During the first visit, where the patient has no prior medical history, direct recommendations are prone to omissions and errors. To address this, \Model~assigns a higher weight (90.43\%) to similar visit information, thereby mitigating these inaccuracies. In the second visit, owing to the continuity of the patient's condition, \Model~can recommend the most suitable medications by relying on historical information, which is assigned a weight of 67.55\%. However, errors in recommendations persist. At this juncture, information from similar visits plays a role in eliminating inappropriate medications. Ultimately, using similar visit information in the model increases the Jaccard similarity of medication recommendations from 46.15\% / 55.17\% to 77.27\% / 64.00\% in Visit-1 / Visit-2. This indicates that information on similar visits can significantly improve the accuracy of recommendations, especially when there is no history of prior medical consultations.

   \begin{table}[htb]
        \centering
        \caption{The example of Patient-908 with two visits for the case study. Medications recommended correctly without using similar visit information are highlighted in \textbf{bold}. \textcolor{blue!60!green}{Blue} and \textcolor{orange!50!red}{\sout{orange}} are used respectively to denote additional correct medications introduced and erroneous medications eliminated after incorporating similar visit information.}
        \begin{tabularx}{0.75\textwidth}{c|X|X}
            \Xhline{1pt}
            &\multicolumn{1}{c|}{{\textbf{Diagnoses}}} & \multicolumn{1}{c}{{\textbf{Medications}}}\\ 
            \Xhline{0.5pt}
            Visit-1 & V8741, V153, 79029, V1083, 5715, 1550, 07054 & N02B, A01A, \textbf{A02B}, \textbf{A06A}, \textbf{B05C}, \textbf{J05A}, \textbf{A12C}, \textbf{A07A}, \textbf{J01E}, \textcolor{blue!60!green}{N01A}, \textbf{C03C}, \textcolor{blue!60!green}{J01C}, \textbf{N02A}, \textbf{B01A}, \textcolor{orange!50!red}{\sout{N05C}}, \textcolor{blue!60!green}{D01A}, \textcolor{orange!50!red}{\sout{J01D}}, \textcolor{orange!50!red}{\sout{L04A}}, \textbf{D11A}, \textbf{A04A}, \textcolor{blue!60!green}{A07E}, \textcolor{blue!60!green}{D07A}, \textcolor{orange!50!red}{\sout{D04A}}, \textcolor{orange!50!red}{\sout{C03D}}\\
            \Xhline{0.5pt}
            Visit-2 & 99682, 78552, 0389, 99592, 5848, 570, 28419, 4561, 5589, 78060, 07070, 5715, V1083, E8780, 4568, V168, V1582 & \textbf{N02B}, \textbf{A01A}, \textbf{A02B}, \textbf{A06A}, \textbf{B05C}, \textbf{A12A}, \textbf{A12C}, \textbf{C01C}, \textbf{A07A}, \textcolor{orange!50!red}{\sout{J02A}}, \textbf{J05A}, \textcolor{orange!50!red}{\sout{V03A}}, \textbf{J01E}, J01C, N02A, \textcolor{orange!50!red}{\sout{A02A}}, \textbf{J01M}, \textbf{B01A}, \textbf{N05C}, \textbf{J01D}, B02B, \textcolor{orange!50!red}{\sout{C08C}}, \textbf{D11A}, A04A, A07E\\
            \Xhline{1pt}
        \end{tabularx}
        \label{tab:case}
    \end{table}



    \section{Conclusion and Future Work}
    \rev{In conclusion, we propose \Model, a hypergraph-based medication recommendation framework that reconstructs a patient’s latent clinical condition by unifying \emph{intra-visit} set-level coherence modeling with \emph{inter-visit} reference augmentation.
    Through a two-stage design, \MedRep\ encodes high-order clinical co-occurrence into a globally consistent, retrieval-friendly embedding space, while \MedRec\ performs visit-conditioned retrieval of similar visits and integrates them with longitudinal history for robust medication prediction.
    Extensive experiments on three real-world benchmarks (MIMIC-III, MIMIC-IV, and eICU) show that \Model\ consistently achieves strong overall performance and outperforms state-of-the-art baselines on key accuracy metrics while maintaining competitive safety.
    Ablation and case studies further verify the complementary contributions of \MedRep\ and \MedRec\ and highlight the benefit of representation--retrieval consistency in utilizing retrieved references.}
    
    \rev{Nevertheless, similar to prior methods~\cite{COGNet, yang2023molerec}, \Model\ may recommend slightly more medications than the ground truth, leading to suboptimal performance on the \#Med.\ metric.
    In addition, embedding-based retrieval can occasionally introduce less relevant visits, especially when the current visit is under-specified.
    In future work, we plan to improve retrieval robustness (e.g., by incorporating stronger relevance filtering or uncertainty-aware retrieval) and enhance prescription compactness to further strengthen the clinical applicability of \Model.}

    \section*{Acknowledgments}
    This research was supported by the Public Computing Cloud of Renmin University of China and by the Fund for Building World-Class Universities (Disciplines) at Renmin University of China.

\normalem
\bibliographystyle{ACM-Reference-Format}
\bibliography{sample-base}

\end{document}